\documentclass{ametsocV6.1_arxiv}
\usepackage{soul}
\usepackage{array}
\usepackage{hyphenat}
\usepackage{gensymb}
\usepackage{siunitx}
\usepackage{macros}
\usepackage{xfrac}
\usepackage{mathtools}
\usepackage{enumitem}
\usepackage{natbib}

\DeclareUnicodeCharacter{03B2}{\ensuremath{\beta}}       

\title{A scaling theory for the macroturbulence of weakly supercritical planetary atmospheres}

\authors{Ryan Eusebi\aff{a}\correspondingauthor{Ryan Eusebi, reusebi@caltech.edu}, Tapio Schneider\aff{a}, Andrew F. Thompson\aff{a}}

\affiliation{\aff{a}{California Institute of Technology, Pasadena, CA}}

\abstract{
A theory for the general circulation of the atmosphere must be based on a theory of its macroturbulence. A central component is the near-surface eddy heat flux, which is closely tied to the mass transport of the surface branch of the circulation. Existing scaling theories are largely formulated within the two-layer quasi-geostrophic framework, which assumes supercritical states and an inverse cascade of kinetic energy---conditions that are frequently not satisfied in planetary atmospheres, including Earth's. Nevertheless, such theories have shown some empirical success, including that the Rhines scale is associated with the mixing length even in the absence of any inverse energy cascade. Here, we analyze hundreds of idealized, dry general circulation model simulations spanning wide ranges of rotation rates, meridional temperature gradients, vertical stratification, and seasonality. Most simulations reside in the marginally critical regime where strongly nonlinear scaling theories fail. We propose a new theory that captures both domain-averaged and local eddy heat flux behavior across all simulations. The theory uses scaling arguments for the zonal momentum balance and accounts for kinematic effects on eddy mixing. The scaling depends on a non-dimensional thermal Rossby number and implicitly accounts for the role of nonlinear eddy momentum fluxes across jets. We demonstrate the skill of the proposed scaling in explaining global near-surface temperature distributions across a wide range of climate states in a simple energy balance model. Implications for general circulation theory and extending the scalings to moist atmospheres are discussed.
}

\begin{document}

\maketitle

%
%
%
\statement
	 The pole-to-equator temperature gradient is a defining characteristic of a planet's climate. Its strength, which regulates mid-latitude weather variability, is determined by a balance between radiative heating and poleward turbulent heat transport by large-scale weather systems. In this work, using hundreds of simulations of planetary atmospheres across a wide spectrum of climate states, we show that classical theories cannot explain the heat transport by these weather systems and derive new scalings that capture the behavior. The scalings are implemented as a turbulence closure in a simple one-dimensional energy balance model to explain the pole-to-equator temperature gradients observed across the simulations. The results offer a new framework for understanding the large-scale turbulence of the atmosphere.

\section{Introduction}

A complete understanding of the time- and zonal-mean state of a planetary atmosphere, given forcing conditions, relies on understanding the emergent macroturbulence of the system. Such macroturbulence consists of the synoptic scale weather systems, pervasive on Earth, that are responsible for the stormy, variable conditions characteristic of the mid-latitudes. These weather systems, or eddies, are a form of large-scale turbulence responsible for the bulk of atmospheric energy transport poleward of the tropics. Macroturbulence is primarily a product of baroclinic instability, through which eddies convert potential energy from horizontal temperature gradients into kinetic energy \citep{lorenz_available_1955}. Heat fluxes associated with these eddies control the meridional temperature gradient of the mid-latitudes; eddy momentum fluxes affect the strength and location of the jet streams, surface winds, and the strength of the Hadley circulation \citep{jeffreys_h_dynamics_1926, walker_eddy_2006, schneider_general_2006}. A long-standing goal of the atmospheric community is a description of eddy fluxes based on mean state conditions. While turbulence is a famously difficult problem to understand, the scale separation between the domain size and the eddies in rapidly rotating planets, the relatively weak nonlinearity of the turbulence \citep{schneider_self-organization_2006}, and the fact that the central goal is understanding spatially inhomogeneous and hence non-trivial mean turbulent fluxes (second moments) rather than an entire hierarchy of higher-order moments, makes understanding their macroturbulence a more tractable problem \citep{held_macroturbulence_1999, vallis_turbulence_2021}.

The near-surface eddy heat flux, which is the focus of this paper, is a particularly important aspect of atmospheric macroturbulence. Its importance is best appreciated by consideration of the equations of motion in isentropic coordinates, in which the atmosphere may conceptually be decomposed into a surface layer and an interior layer. Various studies have shown that the equatorward mass transport in the surface layer is proportional to the near-surface eddy heat flux \citep{andrews_planetary_1976, held_macroturbulence_1999, held_surface_1999, koh_isentropic_2004, schneider_zonal_2005}. The poleward mass transport in the interior of the atmosphere above the surface layer is proportional to the eddy potential vorticity (PV) flux \citep{held_surface_1999, held_macroturbulence_1999}. Since the mass transports of the two layers must balance, a closed theory for either the near-surface eddy heat flux or the interior eddy PV flux could form the basis of a complete general circulation theory. \citet{held_macroturbulence_1999} argues that the near-surface eddy heat flux is a more suitable target because its dynamics more convincingly support a downgradient diffusive representation. Beyond its role in general circulation theory, the near-surface eddy heat flux is widely used in simplified diffusive 1D energy balance models (EBM) to emulate planetary surface climates and climate feedbacks. In these models, the eddy heat flux is crudely represented with a constant diffusivity parameter; they would benefit from a physically motivated parameterization tied to the mean state \citep{held_macroturbulence_1999, chang_role_2023}.

Theories for eddy flux parameterizations have primarily been developed in the quasi-geostrophic (QG) framework, often using 2-layer models on an $f$- or $\beta$-plane with homogeneous forcing conditions \citep{salmon_two-layer_1978, haidvogel_homogeneous_1980, held_quasigeostrophic_1992, panetta_zonal_1993, larichev_eddy_1995, held_scaling_1996, lapeyre_diffusivity_2003, thompson_scaling_2006, thompson_two-layer_2007, arbic_cascade_2007, chang_control_2019, gallet_vortex_2020, gallet_quantitative_2021}. In these QG simulations, the mean vertical shear (and thus, the meridional temperature gradient) and the stratification are imposed. Thus, the criticality of the flow is also externally imposed \citep{held_vertical_1978, schneider_tropopause_2004, schneider_self-organization_2006}. Early theories view the turbulence in terms of energy cascades in baroclinic and barotropic modes, whereby baroclinic potential energy is converted to eddy kinetic energy at the deformation scale from which it cascades to larger scales through the barotropic mode \citep{salmon_two-layer_1978, haidvogel_homogeneous_1980, larichev_eddy_1995, held_scaling_1996, lapeyre_diffusivity_2003}. In these theories, the mixing length is typically set by the Rhines scale \citep{rhines_waves_1975} and the eddy velocity scale depends on the criticality. Later theories, such as the vortex gas scaling theory, have offered an alternative view that recognizes the role of bottom drag \citep{thompson_scaling_2006,thompson_two-layer_2007, gallet_vortex_2020, gallet_quantitative_2021}.

While existing theories explain much of the observed behavior in 2-layer QG systems, the application of these theories to more realistic settings, such as those described by the primitive equations on rotating spheres for atmospheres or oceans, is questionable (see section 3). In QG, the stratification and mean shear are externally imposed, whereas in reality, the stratification and shear are emergent features of the system's equilibration to the external forcing through macroturbulent processes. As suggested by \citet{schneider_tropopause_2004} and \citet{schneider_self-organization_2006}, this limitation of QG ignores the possibility of eddy-mean flow interactions that, for instance, might inhibit the nonlinear eddy-eddy interactions that support cross-scale energy transfer. Similarly, instability in the 2-layer model \citep{phillips_energy_1954} requires that the criticality, a non-dimensional measure of the vertical scale of baroclinic eddies relative to the tropopause depth, be greater than 1, which implies the eddies extend from the bottom layer into the top layer. However, in a continuously stratified system, no such criterion exists as the eddies can set their own dynamical vertical scale \citep{charney_dynamics_1947,held_vertical_1978, vallis_atmospheric_2017}. In fact, while there are exceptions \citep{zurita-gotor_sensitivity_2008,zurita-gotor_circulation_2010, jansen_macroturbulent_2012, jansen_equilibration_2013, jansen_vertical_2013, chai_role_2014, chemke_latitudinal_2015}, idealized simulations across a wide parameter regime indicate that planetary atmospheres tend to reside in marginally critical states and exhibit no inverse energy cascade \citep{schneider_self-organization_2006, schneider_scaling_2008}. Earth's atmosphere likewise exhibits no inertial range with a substantial inverse energy cascade \citep{boer_large-scale_1983, schneider_self-organization_2006}. Additionally, the assumptions of QG might break down in the surface layer of the atmosphere, for instance, at fronts with steep near-surface temperature gradients \citep{held_surface_1999, schneider_boundary_2003, schneider_zonal_2005}. Finally, many QG theories have been developed in the homogeneously forced setting, which does not include eddy momentum flux convergences and intense zonal jets that can modulate eddy behaviors\citep{greenslade_vertical_2008, schneider_scaling_2008, ferrari_suppression_2010}, particularly in the nonlinear decay stage of the eddy life cycle \citep[e.g.,][]{simmons_life_1978, james_suppression_1987, nakamura_midwinter_1992}.

Despite these caveats, QG theories have been applied extensively with mixed results in more realistic settings, such as primitive equation models and comprehensive GCMs. \citet{jansen_macroturbulent_2012, jansen_equilibration_2013} apply the \citet{held_scaling_1996} scalings with empirical success for supercritical Boussinesq channel simulations. However, \citet{jansen_macroturbulent_2012} notes a degradation in the scaling for marginally critical simulations. \citet{barry_poleward_2002} similarly invoke the inverse cascade scalings, and demonstrate a close relation between the Rhines scale and the mixing length in a series of comprehensive GCM simulations with forcing conditions similar to Earth's. \citet{chang_scaling_2022} apply the scalings from \citet{lapeyre_diffusivity_2003} and \citet{barry_poleward_2002} to explain domain-averaged eddy diffusivities for vertically integrated eddy heat fluxes in GCM simulations. Other studies have shown that kinematic effects, associated with a suppression in eddy mixing when eddy phase speeds and the mean zonal flow differ, impact eddy mixing rates \citep{ferrari_suppression_2010, srinivasan_reynolds_2014}; this process has traditionally been neglected in classical scalings. \citet{schneider_scaling_2008}, who deviate from the classic QG theories, introduce bulk scaling laws that accurately describe domain-averaged atmospheric macroturbulence across a spectrum of climates; however, the scalings exhibit an unexplained power law dependence on rotation rate and planetary radius and depend on the baroclinic zone width. The bulk nature of these scalings and their dependence on baroclinic zone width, which itself is defined in terms of the equilibrated eddy flux statistics, complicate their application as a local closure in an EBM. Overall, no study has demonstrated an accurate local closure valid across a wide regime of planetary atmospheres.

This study introduces new, closed scalings for the near-surface eddy heat fluxes based on the analysis of hundreds of dry GCM simulations spanning a broad range of rotation rates, meridional temperature gradients, seasonality, and convective lapse rates. Many of these simulations, which are very similar to those of \citet{schneider_self-organization_2006} and \citet{schneider_scaling_2008}, reside in the marginally critical regime, like Earth, where classical scalings fail. The new theory amounts to a diffusive closure that depends on an eddy mixing length, an eddy velocity scale, and a correlation coefficient. The theory directly accounts for suppressive mean flow effects \citep{ferrari_suppression_2010} on the mixing length and correlation coefficient and uses scaling arguments on the non-dimensional zonal momentum balance to determine the characteristic eddy velocity scale. Importantly, the results depend not on the criticality parameter, which QG theories suggest, but instead on a non-dimensional number akin to the thermal Rossby number \citep{held_nonlinear_1980}. 

The rest of the paper is organized as follows: section 2 describes the idealized GCM, along with diagnostic results and calculation conventions. Section 3 develops the scaling theory. In section 4 we demonstrate the skill of the proposed heat flux closure by implementing the new eddy diffusivity in a simple 1D EBM and accurately reproducing global surface temperature distributions across a wide range of climates. Section 5 discusses implications for general circulation theory and section 6 presents the conclusions.

\section{Idealized GCM simulations and diagnostics}

\subsection{Model details}

We use the idealized GCM from \citet{schneider_tropopause_2004} and \citet{schneider_self-organization_2006} with differences in the configuration described below. The GCM, with modifications, has been used in various idealized studies concerning the general circulation \citep[e.g.,][]{walker_eddy_2006, schneider_eddy-mediated_2008, vallis_isca_2018}. It is a primitive equation, spectral-transform model built on the Flexible Modeling System (FMS) available from the Geophysical Fluid Dynamics Laboratory (GFDL). The dynamical core is identical to the spectral dynamical core described in \citet{held_proposal_1994}. The horizontal resolution is varied between T42, T85, T127, and T213 depending on rotation rate (see Table \ref{tab:param_sweep}). The vertical discretization uses a centered difference scheme with 30 vertical levels in a $\sigma$-coordinate system \citep[chapter 7.6 of][]{durran_numerical_1999, schneider_tropopause_2004} with $\sigma=p/p_s$ for the instantaneous surface pressure $p_s$. All simulations are run to steady-state, with a spin-up period of at least 200 days and eddy statistics calculated over a subsequent period of at least 200 days. No significant changes to the eddy statistics are observed with longer averaging or spinup periods. 

As in \citet{held_proposal_1994}, diabatic processes are represented by Newtonian relaxation towards a prescribed radiative equilibrium (RE) profile. Unlike \citet{held_proposal_1994}, a statically unstable RE profile is used \citep{schneider_tropopause_2004}. Such an equilibrium profile is more realistic, and in conjunction with a dry convection scheme, allows the atmosphere to equilibrate with an emergent combination of the stabilizing effects of both convection and dynamical heat transport. We use the RE profile described in \citet{eusebi_theory_2026}, which is similar to that used in \citet{schneider_eddy-mediated_2008}. The surface RE temperature profile is described by
\begin{equation}
\label{eq:surfacere}
    T_s^e = T_s^\mathrm{avg} - \Delta_h\left(  \sin^2\phi - 2\sin\phi\sin\phi_0 - \frac{1}{3}\right).
\end{equation}
Here, $T_s^\mathrm{avg}$ is the area-weighted global average surface temperature, $\Delta_h$ is the surface RE pole-to-equator surface temperature contrast, and $\phi_0$ is the latitude of maximum $T_s^e$. As in \citet{schneider_tropopause_2004}, the vertical structure of the RE profile is described by
\begin{equation*}
    T^e(\phi, p) = T_t^e\left[1 + d_0(\phi)\left(\frac{p}{p_0}\right)^\alpha \right]^{1/4},~~~~~d_0 = \left(\frac{T_s^e(\phi)}{T_t^e}\right)^4 - 1.
\end{equation*}
Here, $T_t^e$ is the skin temperature at the top of the atmosphere (set to 200 K), $p_0$ is the reference surface pressure (set to 1000~hPa), and the exponent $\alpha=3.5$ controls the lapse rate of the RE state. This vertical profile is a solution of the two-stream equations for a semigray atmosphere \citep{schneider_tropopause_2004}. Temperatures are relaxed toward the RE profile according to a spatially homogeneous radiative timescale $\tau_r$, which is varied across experiments as described in the next section. 

The simulations do not explicitly represent moisture. However, a dry convection scheme, mimicking the stabilizing effect of latent heat release from moist convection, relaxes the atmosphere to a rescaled dry adiabat $\gamma\Gamma_d$ (dry adiabatic lapse rate $\Gamma_d$, rescaling parameter $\gamma\le 1$) over a timescale of $\tau_c=1~\mathrm{day}$ whenever lapse rates exceed the convective lapse rate \citep{schneider_tropopause_2004}. The scheme operates globally, as in \citet{walker_eddy_2006} and \citet{schneider_eddy-mediated_2008}.

Boundary layer turbulence that mimics the effect of turbulent dissipation is represented as quadratic drag with an inverse length scale drag coefficient $\mu(\sigma)$ that varies linearly from $\mu_s$ at the surface to 0 at $\sigma_b=0.85$, the top of the boundary layer \citep{schneider_tropopause_2004}. We choose the value $\mu_s = \SI{5e-6}{\per\m}$ which is common in the literature \citep[e.g.,][]{schneider_tropopause_2004, schneider_self-organization_2006} and realistic for terrestrial atmospheric boundary layers. The results presented are not sensitive to modest changes of the bottom drag. Biharmonic hyperdiffusion with a coefficient of $\SI{6e-5}{\m^4~\s^{-1}}$ is included to damp small-scale instabilities. 

\subsection{Simulations across a wide range of parameters}

The simulations survey a wide parameter regime of planetary atmospheres and forcing conditions. Rotation rates $\Omega_* = \Omega/\Omega_E$ are varied between 0.5 and 4 times Earth's rotation rate $\Omega_E$; equator-to-pole RE temperature differences $\Delta_h$ vary between 40 and 200~K; $\phi_0$ (a metric of seasonality) varies between 0 and 20$\degree$; the convective lapse rate rescaling parameter $\gamma$ varies between 0.7 and 1.0; and the radiative relaxation timescale $\tau_r$ varies between 30 and 60 days. The exact parameter values are summarized in Table \ref{tab:param_sweep}. Every combination of $\Omega_*$, $\Delta_h$, and $\gamma$ was run with $\phi_0=0\degree$ and $\tau_r=30$ days. For the remaining values of $\phi_0$ and $\tau_r$, each unique value was run with every combination of $\Omega_*$ and $\Delta_h$ with the default values of $\phi_0=0\degree$ (if $\phi_0$ is not varied), $\tau_r=30$~days (if $\tau_r$ is not varied), and $\gamma=0.7$. The $\Omega_*=4$ and 2 simulations with $\gamma=1.0$ and $\Delta_h=40~$K are omitted from the results because vanishingly small static stabilities complicated their analysis. In total, this yields 323 GCM simulations across a broad range of climate states. Such a large ensemble, like in \citet{schneider_scaling_2008} and \citet{levine_baroclinic_2015}, is necessary to fully test the generality of any theory. We are unaware of a previous study that has included seasonality as a variable parameter in a comprehensive study of planetary atmosphere eddy fluxes.  \citet{jansen_macroturbulent_2012, jansen_equilibration_2013} show that the planetary scale $f/\beta = a\tan\phi$ affects criticality and eddy flux behavior in a baroclinic zone; this scale is set by their channel simulation configuration, since $f$ and $\beta$ are prescribed. However, on a rotating sphere, $f/\beta$ is set by the latitudinal band over which the equilibrated baroclinic zone occurs. All else equal, varying $\phi_0$ can change the latitude of the equilibrated baroclinic zone due to variations in the thermal forcing and interactions with the Hadley circulation.

\begin{table}[]
    \centering
    \begin{tabular}{l|l}
        Parameter & Values \\
        \hline 
        \hline
        $\Omega_*$ & 0.5, 1, 2, 4 \\
        $\Delta_h$ (K) & 40, 60, 80, 100, 120, 140, 160, 180, 200 \\
        $\phi_0$ ($\degree$) & 0, 5, 10, 15, 20 \\ 
        $\gamma$ & 0.7, 0.8, 0.9, 1.0 \\
        $\tau_r$ (days) & 30, 60
    \end{tabular}
    \caption{Parameter space of simulations explored in this study. Simulations are run for various combinations of these parameters, as described in the main text.}
    \label{tab:param_sweep}
\end{table}

\subsection{Simulation eddy energies and criticalities}

\begin{figure}[t]
    \centering

    \includegraphics[width=1.0\linewidth]{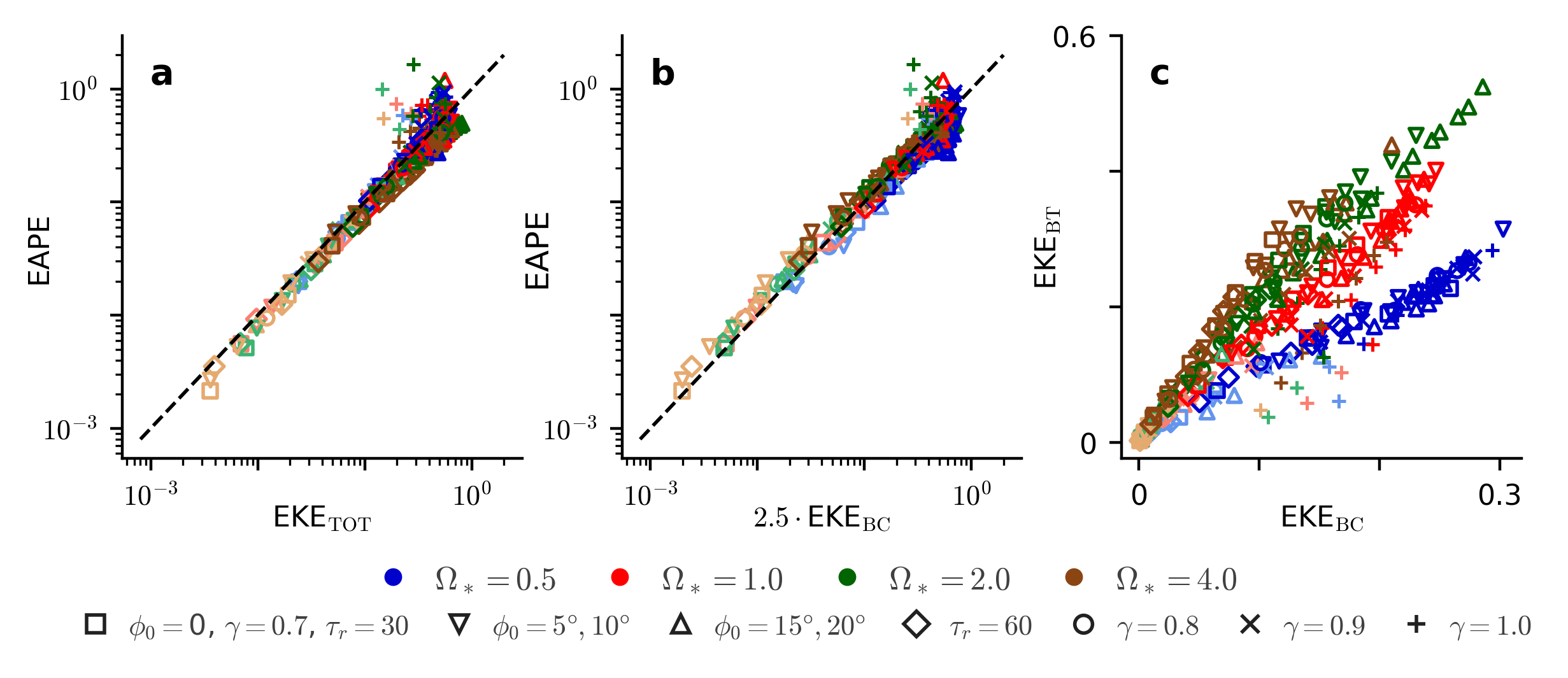}
    \caption{Eddy energy statistics per unit area averaged over the winter hemisphere from each simulation, with each marker representing the statistics from one simulation. The color indicates $\Omega_*$ and the shape indicates $\phi_0$, as indicated in the legend. Low $\phi_0$ indicates $\phi_0\leq 5\degree$, and high $\phi_0$ indicates $\phi_0 > 5\degree$. Lighter color shadings for a given color represent $\Delta_h < 80~$K and darker shadings indicate $\Delta_h\geq 80~$K. The units for all eddy energies are MJ~m$^{-2}$. (a) EAPE vs. EKE$_\mathrm{TOT}$. (b) EAPE vs. EKE$_\mathrm{BC}$. The dashed line indicates the relation $\mathrm{EAPE} = 2.25~\mathrm{EKE}_\mathrm{BC}$. (c) EKE$_\mathrm{BT}$ vs. EKE$_\mathrm{BC}$.}
    \label{fig:eddy_energies}
\end{figure}

Eddy energy statistics, averaged over the winter hemisphere, summarize the range of atmospheric behaviors represented in the 323 simulations (Figure \ref{fig:eddy_energies}). For almost all simulations, energy is equipartitioned between eddy available potential energy (EAPE) and EKE, except for some of the $\gamma=1.0$ simulations (Figure \ref{fig:eddy_energies}a). Here, EAPE is defined using Lorenz's approximate definition in terms of potential temperature variations on pressure surfaces \citep{lorenz_available_1955} vs. the total eddy kinetic energy $\mathrm{EKE}_\mathrm{TOT}$. The dashed line indicates the relation $\mathrm{EAPE} = \mathrm{EKE}_\mathrm{TOT}$. The EAPE and $\mathrm{EKE}_\mathrm{BC}$ exhibit a linear relationship (Figure \ref{fig:eddy_energies}b) with a proportionality factor of 2.25, which is consistent with \citet{schneider_self-organization_2006} and similar to the expected equipartitioning for the linearly most unstable waves. If there was an inverse cascade of energy to large scales in the simulations, the EAPE would scale with the barotropic EKE ($\mathrm{EKE}_\mathrm{BT}$), which would be much greater than $\mathrm{EKE}_\mathrm{BC}$ \citep{held_scaling_1996, schneider_self-organization_2006}. This relationship is consistent with the absence of an inverse energy cascade, which is common in simulations of planetary atmospheres \citep{schneider_tropopause_2004, schneider_self-organization_2006}. The exceptions include simulations with low $\Delta_h$ and $\gamma=1.0$, in which the deformation radii become exceedingly small, criticalities become large, and there is spectral room for a modest inverse cascade to operate. Additionally in these simulations, the vanishingly small static stabilities can introduce errors in the calculation of the EAPE. The $\mathrm{EKE}_\mathrm{BT}$ and $\mathrm{EKE}_\mathrm{BC}$ maintain a remarkably linear relationship across many simulations, but the linear fit varies across different rotation rates (Figure \ref{fig:eddy_energies}c) consistent with \citet{schneider_scaling_2008}. Faster $\Omega_*$ simulations exhibit a greater proportion of EKE in the barotropic mode. 

\begin{figure}[t]
    \centering
    \includegraphics[width=0.45\linewidth]{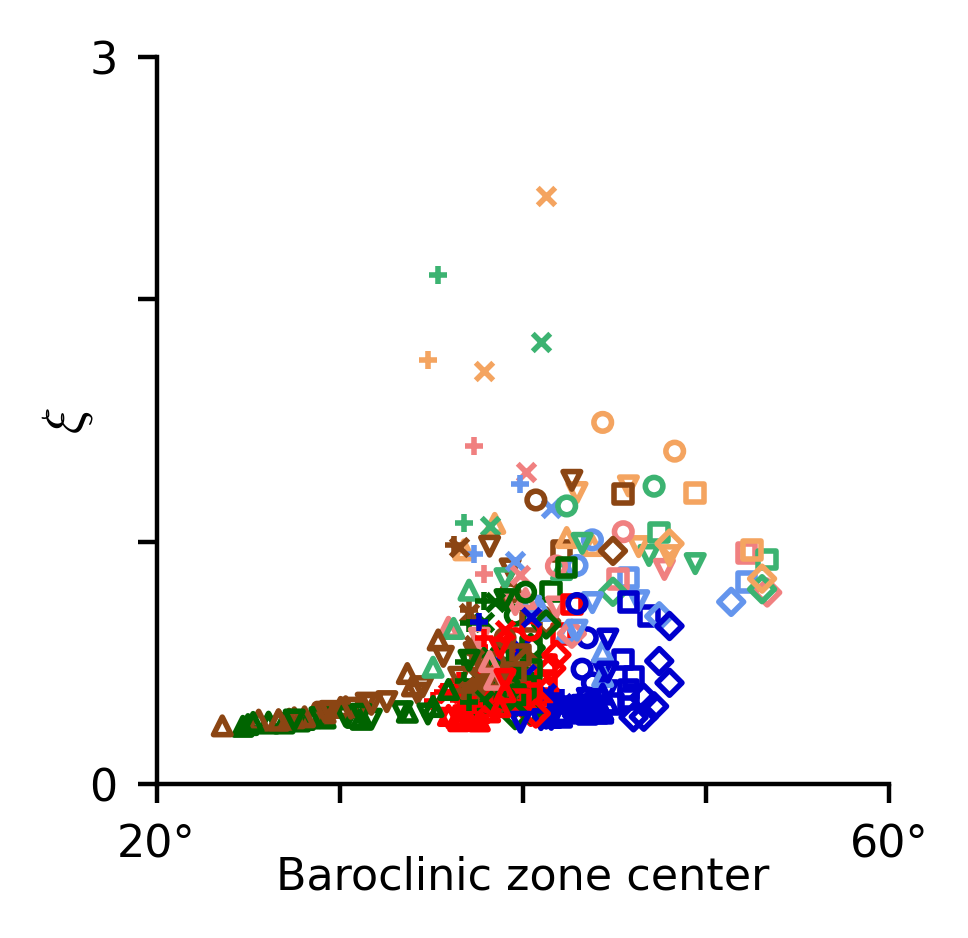}
    \caption{Criticality $\xi$ calculated with (\ref{eq:criticality}) averaged over the baroclinic zone vs. the latitude of the centroid of the baroclinic zone.}
    \label{fig:criticality}
\end{figure}

Analysis of the baroclinic zone-averaged criticality $\xi$ indicates that the vast majority of the simulations reside in a marginally critical or subcritical state (Figure \ref{fig:criticality}), consistent with the eddy energies in Fig. \ref{fig:eddy_energies}. Only some low-$\Delta_h$ simulations with high $\gamma$ attain modestly larger $\xi$. Here, $\xi$ is approximated in terms of its near-surface quantity (similar to \citet{jansen_macroturbulent_2012}) as:
\begin{equation}
\label{eq:criticality}
    \xi = \frac{f}{\beta} \frac{a^{-1}\partial_\phi{\overline{\theta}}_s}{H_t\overline{\partial_z\theta}^s}.
\end{equation}
The planetary radius $a$ is equal to Earth's radius for all simulations. The Coriolis parameter is indicated by $f$ and its meridional derivative by $\beta$. Across all simulations, we set $H_t=10~$km. While the tropopause height does vary slightly across simulations, its variations are small compared to all other quantities in our analyses so the approximation only minimally affects the results. The notation $\ov{(\cdot)}$ represents a time and zonal mean along a pressure surface, $\ov{(\cdot)}^s$ indicates an average evaluated over the near-surface pressure range, and the subscript $(\cdot )_s$ indicates a near-surface-averaged quantity.\footnote{For simplicity, all averages and derivatives are shown as being calculated on surfaces of constant pressure $p$ or log-pressure $z$. Note, however, that the model output is on $\sigma$-levels, so all derivatives are along $\sigma$-levels and all averages are surface-pressure weighted averages along constant $\sigma$-levels, as in \citet{schneider_self-organization_2006}. The near-surface level refers to the sigma-coordinates between 1.0 and 0.7, which corresponds to the pressure definition specified in the main text.} All near-surface quantities are calculated as averages between 700 hPa and 1000 hPa. This pressure range corresponds to the surface layer in isentropic coordinates referred to in \citet{held_surface_1999}. For the remainder of the manuscript, the baroclinic zone refers to all latitudes in the winter hemisphere where the near-surface eddy heat flux is greater than 50\% of the simulation maximum. 

\subsection{The eddy heat flux and diffusivity}

\begin{figure}
    \centering
    \includegraphics[width=0.5\linewidth]{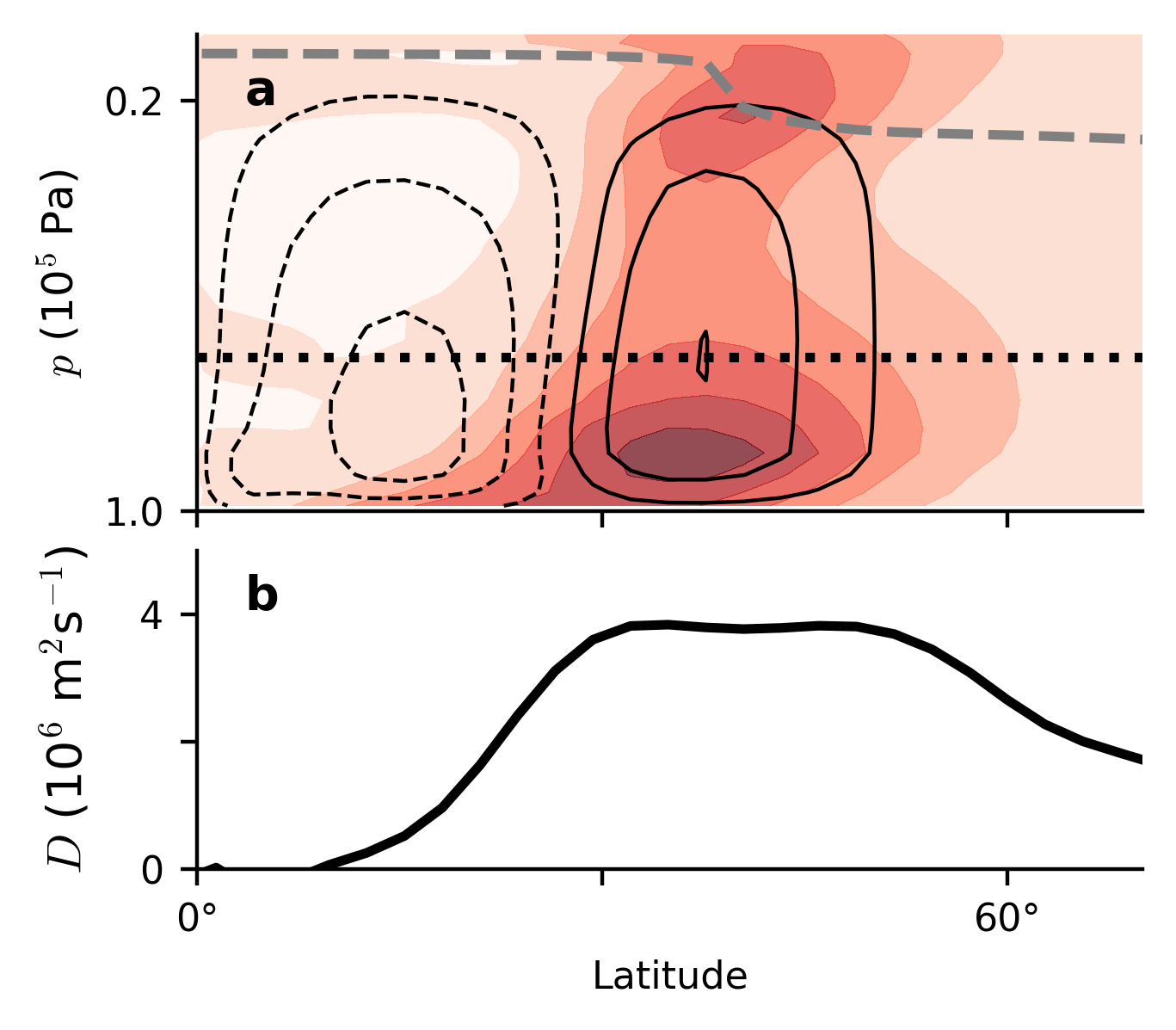}
    \caption{GCM results for an Earth-like simulation with $\phi_0=0\degree$, $\Delta_h=120~$K, $\Omega_*=1$, and $\tau_r=30~$d. (a) The zonal- and time-mean latitude-pressure profile of the eddy heat flux $\overline{v'\theta'}^s\cos\phi$, indicated by the shading with contour intervals of \SI{5}{\K\m\per\s}, with the darkest shading indicating greater than \SI{25}{\K\m\per\s}. Black contour lines indicate mean meridional circulation streamfunction contours (solid for counter-clockwise and dashed for clockwise) with contour intervals of \SI{30e9}{\kg\per\s}. The thick gray dashed line indicates the tropopause, defined by the lowest altitude at which the vertical lapse rate reaches $\SI{2}{\K\per\km}$. The thick dotted black line indicates the near-surface pressure level which is the focus of this work. (b) The near-surface diffusivity $D$ calculated with (\ref{eq:heatflux_D}). }
    \label{fig:heat_D_profile}
\end{figure}

Figure \ref{fig:heat_D_profile}a shows an example of the zonal and temporally averaged eddy heat flux profile for an Earth-like simulation. The upper boundary of our near-surface pressure range is indicated by the thick black dotted line. Note that the majority of the eddy heat fluxes tend to be confined to the near-surface layer. It is useful to understand these eddy heat fluxes in terms of the diffusivity $D$, defined through a downgradient closure as:
\begin{equation}
\label{eq:heatflux_D}
    \overline{v'\theta'}^s = -D\frac{1}{a}\fpd{\overline{\theta}_s}{\phi}.
\end{equation}
In Earth-like simulations, the near-surface diffusivity $D$ varies with latitude and attains its largest value in the vicinity of the strongest heat fluxes (Figure \ref{fig:heat_D_profile}b). The remainder of this work will aim to understand what controls $D$. 

Following \citet{thompson_scaling_2006}, $D$ can be expressed in terms of an eddy mixing length $\ell_m$ and an eddy velocity scale $v_e$:
\begin{equation}
\label{eq:diff_closure}
    D = c v_e\ell_m.
\end{equation}
The eddy velocity scale $v_e$ is defined as the near-surface rms velocity fluctuations:
\begin{equation*}
    v_e = \left(\overline{v'^2}^s\right)^{1/2}.
\end{equation*}
The mixing length $\ell_m$ is defined from the rms temperature fluctuations and the mean temperature gradient: 
\begin{equation}
\label{eq:mixl_definition}
    \ell_m = \frac{\left(\overline{\theta'^2}^s\right)^{\frac{1}{2}}}{a^{-1}\partial_\phi \overline{\theta}_s}.
\end{equation}
The correlation coefficient $c$ is defined as:
\begin{equation}
\label{eq:correlation}
     c = \frac{~\overline{v'\theta'}^s}{\sqrt{\overline{v'^2}^s~\overline{\theta'^2}^s}}.
\end{equation}
The correlation $c$ is assumed constant in many QG studies \citep{thompson_scaling_2006, thompson_two-layer_2007}. However, in this study $c$ varies appreciably between different climate states and latitudinally within a given climate state (section 3c). Its effects must be accounted for to obtain an accurate scaling. The remainder of this manuscript will develop scalings for $\ell_m$, $v_e$, and $c$ to obtain a fully closed scaling for $D$. 

\section{Theory}

We seek scalings for the equilibrated near-surface eddy heat flux diffusivity $D$, which unlike 2-layer QG theories, does not require \textit{a priori} assumptions of the criticality, the existence of an inverse energy cascade, and the partitioning of the eddy kinetic energy between the barotropic and baroclinic modes. The eddy velocity scale $v_e$ is simply its near-surface value. To make contact with previous theories, we will often focus on non-dimensional quantities. The non-dimensional eddy velocity is given by $v_e^* = v_e/U$, where $U$ in primitive equations is defined as the thermal wind shear over the depth of the troposphere:
\begin{equation}
\label{eq:U_primitive}
    U = H_t\fpd{\ov{u}}{z} = \frac{gH_t}{f\ov{\theta}_s a}\fpd{\overline{\theta}_s}{\phi}.
\end{equation}
We calculate $U$ in terms of $\partial_\phi \overline{\theta}_s$ with applications to 1D EBMs in mind. The non-dimensional eddy mixing length is given by $\ell_m^* = \ell_m/\lambda$, with the outer Rossby deformation radius calculated as:
\begin{equation}
\label{eq:rossby_outer}
    \lambda = \frac{NH_t}{f},
\end{equation}
where $N^2 = \frac{g}{\overline{\theta}_s} \overline{\partial_z \theta}^s$ is defined by its near-surface quantity for simplicity. Note that with these definitions, it follows that the criticality in (\ref{eq:criticality}) can be expressed as $\xi=U/\beta \lambda^2$.

\subsection{The mixing length and its suppression by the mean flow}

\begin{figure}
    \centering
    \includegraphics[width=1.0\linewidth]{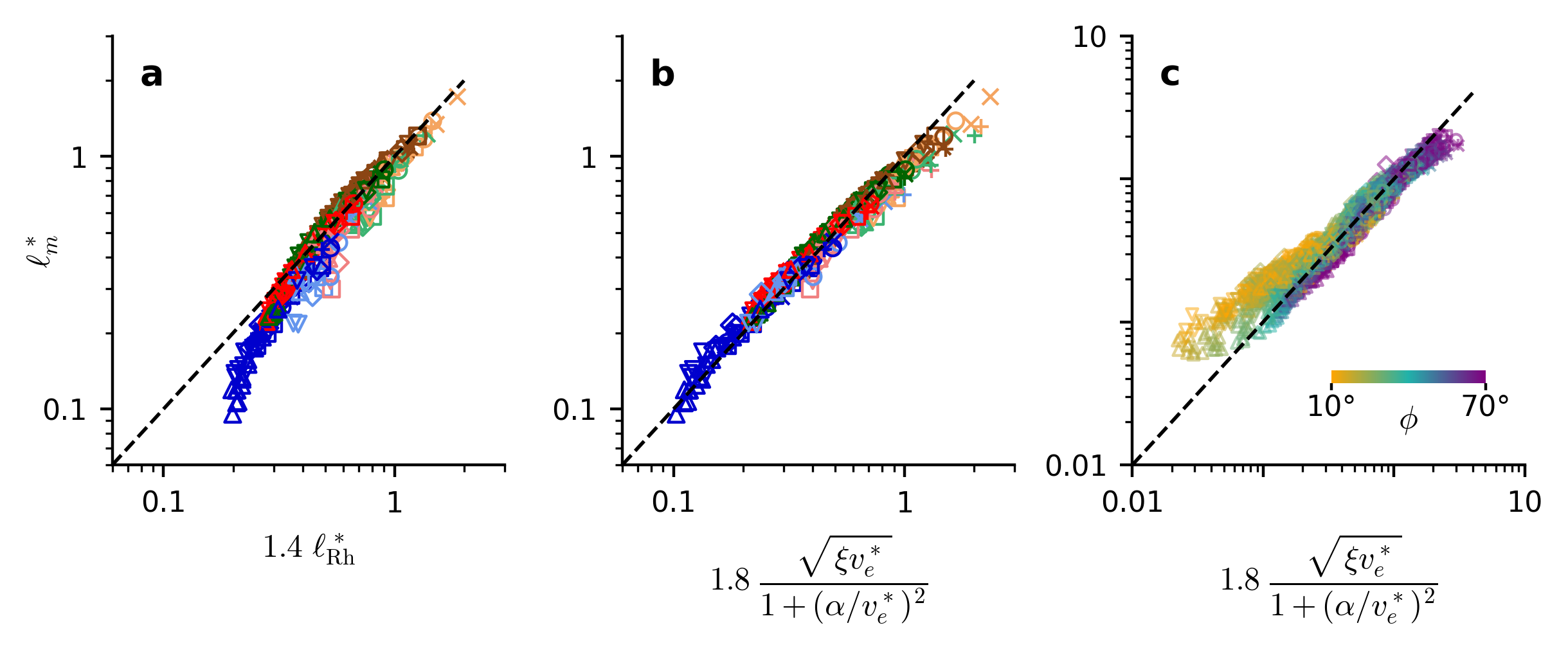}
    \caption{Scalings for the non-dimensional mixing length $\ell_m^*$ with empirically determined scaling coefficients indicated in the x-axis label. (a) The Rhines scale averaged over the baroclinic zone in each simulation. Marker indicators are as in Fig. \ref{fig:eddy_energies}. (b) Like a, but with the full scaling (\ref{eq:mixl_supp_star}) that accounts for the suppression by the mean flow. Note that $\alpha=0.1$. (c) Like b, but showing the scaling evaluated at local point-wise estimates in the winter hemisphere from all the simulations in b. The color represents the latitudinal band of the point, as indicated in the colorbar on the axis. Ten points are plotted from each simulation, evenly spaced between the highest and lowest latitudes in the winter hemisphere where the near-surface eddy heat flux is greater than 20\% of the winter hemisphere maximum in the simulation. Over 3,000 total points are plotted in this panel.}
    \label{fig:lmix}
\end{figure}

A popular choice for the mixing length is the Rhines scale:
\begin{equation}
    \ell_\mathrm{Rh} = \sqrt{\mathrm{EKE}_{\mathrm{BT}}^{1/2}/\beta},
\end{equation}
as suggested by \citet{held_scaling_1996, barry_poleward_2002, chang_scaling_2022}. For slower rotation rates,  $\ell_\mathrm{Rh}$ systematically overestimates $\ell_m$; also, biases persist between different rotation rates (Figure \ref{fig:lmix}a). The scaling only provides a good fit if the planetary parameters are not varied sufficiently \citep[particularly the rotation rate, as in][]{barry_poleward_2002}. To rectify these errors, we consider the kinematic effects that mean flow advection of potential temperature anomalies will have on $\ell_m$.

Following other studies \citep{chang_scaling_2022}, dimensional considerations suggest that the characteristic decorrelation timescale for eddy mixing should depend on the eddy velocity scale $v_e$ and $\beta$ through:
\begin{equation}
\label{eq:gamma_beta}
    \gamma_\beta = \sqrt{v_e\beta},
\end{equation}
where $\gamma_\beta$ is an inverse decorrelation timescale. The product $v_e \gamma_\beta^{-1}$ yields the length scale:
\begin{equation}
\label{eq:l_beta}
    \ell_\beta = \sqrt{\frac{v_e}{\beta}}.
\end{equation}
This is similar to the Rhines scale, except that the length scale is expressed in terms of the near-surface eddy velocity $v_e$, not the barotropic eddy velocity. Unlike other studies, we do not assume that the barotropic component dominates the heat flux (as it does in 2-layer QG, where the baroclinic mode is itself equivalent to the temperature anomalies). In fact, in the simulations the contribution of the baroclinic and barotropic modes to the total eddy heat flux is proportional to the EKE in that mode (not shown). As such, the near-surface velocity $v_e$ is the relevant velocity scale.

Various studies identify a suppression of the meridional eddy diffusivity when the zonal mean flow differs from the zonal eddy phase speed \citep{ferrari_suppression_2010, klocker_estimating_2012, srinivasan_reynolds_2014}. Under these conditions, tracer filaments may be peeled out of an eddy by the mean zonal flow before substantial meridional mixing can occur, which reduces the mixing length and suppresses the diffusivity. Accounting for suppressive effects and following \citet{ferrari_suppression_2010}, the mixing length is expressed in terms of the characteristic eddy decorrelation timescale $\gamma^{-1}$ and velocity scale $v_e$ as:
\begin{equation}
\label{eq:mixl_supp}
\ell_m = \frac{k^2}{\kappa^2}\frac{\gamma}{\gamma^2 + k^2[c_w - \overline{u}(p)]^2}v_e,
\end{equation}
where $k$ and $l$ are the zonal and meridional wavenumber of the eddy respectively, $\kappa^2 = k^2 + l^2$, $c_w$ is the phase speed of the eddy, and $\overline{u}(p)$ is the mean zonal velocity at pressure $p$. First, assume the wavenumber ratio $k^2/\kappa^2$ is an O(1) constant. We substitute $\gamma = \gamma_\beta$ and assume that the characteristic eddy scale is set by similar dimensional reasoning such that $k\sim \ell_\beta^{-1}$. To understand the eddy phase speed $c_w$, consider that a precursor for baroclinic instability is phase-locking of two Doppler-shifted counter-propagating Rossby waves in the vertical direction. The phase speed at which this phase-locking occurs will take on a value characteristic of the mid-troposphere mean zonal flow \citep{ferrari_suppression_2010}. Consequently, the eddy phase speeds of the baroclinically unstable upper and lower tropospheric Rossby waves will differ from the upper and lower tropospheric mean zonal wind, assuming a non-zero vertical shear $U$. Therefore, at the near-surface pressure level, we can express the difference between the phase speed and mean near-surface zonal flow $\overline{u}_s$ as some fraction $\alpha$ of the mean shear $U$:
\begin{equation}
    c_w - \overline{u}_s \approx \alpha U.    
\end{equation}
Plugging in these quantities into (\ref{eq:mixl_supp}) and dividing the numerator and denominator by $\gamma_\beta$ yields our mixing length scaling accounting for mean flow suppression:
\begin{equation}
    \ell_m \sim \frac{\ell_\beta}{1 + \alpha^2U^2/v_e^2},
\end{equation}
which in non-dimensional form is expressed as:
\begin{equation}
\label{eq:mixl_supp_star}
    \ell_m^* \sim \frac{\sqrt{\xi v_e^*}}{1 + (\alpha/v_e^*)^2}.
\end{equation}
The suppression factor becomes negligible when $v_e \gg \alpha U$. The Charney model of baroclinic instability \citep{charney_dynamics_1947} suggests steering levels (the vertical level at which the eddy phase speed equals the mean zonal flow) below the mid-troposphere level. Indeed, setting $\alpha=0.1$ in (\ref{eq:mixl_supp_star}), consistent with a steering level around 700~hPa, yields accurate mixing length predictions across the simulations (Figure \ref{fig:lmix}b). The resulting suppression factor is significant and varies in magnitude across all simulations, becoming negligible for high $\Omega_*$ simulations and increasing to as strong as 50\% for some low rotation rate simulations. Overall, the scaling explains 98\% of the variance across all simulations in log-space (Fig. \ref{fig:lmix}b). Figure \ref{fig:lmix}c shows the same scaling (\ref{eq:mixl_supp_star}), but applied locally at various individual latitudes (indicated by the marker color) throughout all simulations shown in panel b. The scaling is able to capture the local mixing length behavior accurately: it explains 95\% of the variability in the non-dimensional mixing length across more than 3,000 points plotted in log-space. The scaling over-predicts the suppression at some low latitude points, where heat fluxes are already weak and induced by surface-trapped waves.

\subsection{The eddy velocity scale}

\citet{schneider_self-organization_2006} and \citet{schneider_scaling_2008} demonstrate that across a wide regime of planetary atmospheres, the vertically and domain-integrated (across the baroclinic zone) EKE scales with the EAPE, which in turn scales with mean available potential energy (MAPE) integrated over the baroclinic zone. However, since the baroclinic zone is itself a feature of the equilibrated turbulence, a fully closed scaling should ideally be independent of this feature. Additionally, these scalings were developed to explain the bulk turbulence properties of atmospheres. For a local scaling, a similar equipartitioning argument can be invoked to relate the local EKE to the MAPE integrated over the characteristic mixing length $\ell_m$ of eddies at a given latitude. However, in Appendix A, we show that the functional form of the mixing length expression (\ref{eq:mixl_supp_star}) complicates using this assumption to derive a closed velocity scaling. Additionally, the GCM results indicate quantitative errors when applying this assumption locally.

Instead, we use scaling arguments on the zonal momentum balance to determine an eddy velocity scale. Consider the time- and zonal-average zonal momentum balance in the upper troposphere:
\begin{equation}
\label{eq:zonal_momentum}
    \frac{\ov{v}}{a}\fpd{\ov{u}}{\phi} + \ov{w}\fpd{\ov{u}}{z} - f\ov{v} = -\frac{1}{a\cos^2\phi}\fpd{}{\phi}\left( \ov{u'v'} \cos^2\phi \right). 
\end{equation}
Here we have neglected a frictional term (friction primarily acts in the boundary layer through the quadratic drag formulation) and the vertical eddy momentum flux convergence term. We scale the variables as $\ov{u}=U\hat{u}$, $\ov{v}=V\hat{v}$, $\ov{w}=VH_t/a\hat{w}$, $z=H_t\hat{z}$, and $\ov{u'v'} = V_e^2 \ov{\hat{u}^*\hat{v}^*}$, where terms with $\hat{(\cdot)}$ are non-dimensional. The scaling for $\ov{w}$ derives from scaling the zonally averaged continuity equation. The zonal and meridional eddy velocities are assumed to scale isotropically as $V_e$. The upper troposphere mean zonal velocity scale $U$ may be expressed in terms of the meridional temperature gradient using the thermal wind balance (\ref{eq:U_primitive}). We define the non-dimensional parameter $\chi$:
\begin{equation}
    \chi = \frac{U}{fa} = \frac{gH_t}{f^2a^2} \frac{\partial_\phi \theta}{\ov{\theta}_s}.
\end{equation}
Dividing through by $fV$ and plugging in the scalings, the zonal momentum balance  (\ref{eq:zonal_momentum}) becomes:
\begin{equation}
\label{eq:zonal_chi}
    \chi \hat{v} \fpd{\hat{u}}{\phi} + \chi \hat{w}\fpd{\hat{u}}{\hat{z}} - \hat{v} = -\frac{V_e^2}{UV}\chi \frac{1}{\cos^2\phi}\fpd{}{\phi}(\ov{\hat{u}^* \hat{v}^*} \cos^2 \phi).
\end{equation}
The parameter $\chi$ is akin to a planetary scale Rossby number comparing the thermal wind shear $U$ against the planetary velocity scale $fa$. For rapdily rotating planets like Earth, $\chi \ll 1$. This number is similar to the ``thermal Rossby number'' defined in \citet{held_nonlinear_1980} for describing axisymmetric Hadley circulations. For $\chi \ll 1$, the eddy momentum flux divergence (EMFD) term must balance the Coriolis term at leading order. For the EMFD term to be O(1), it must be true that the non-dimensional upper tropospheric eddy velocity scales as:
\begin{equation}
    \frac{V_e^2}{U^2} \sim \frac{V}{U}\frac{1}{\chi}.
\end{equation}
In QG theories, the ageostrophic meridional velocity $V$ is assumed to scale with $U$ multiplied by the Rossby number $\ro$. Given $\chi$ is similar to $\ro$ for the planetary scale primitive equations, it is tempting to assume $V\sim U\chi$. Doing so renders the non-dimensional eddy velocity $V_e/U$ a constant, yet the non-dimensional eddy velocity varies by an order of magnitude across all simulations. 

Instead, consider that the zonal momentum balance of the upper troposphere (\ref{eq:zonal_chi}) can be expressed in terms of just three non-dimensional numbers $\chi$, $V_e/U$, and $V/U$. The latter number $V/U$ does not appear at leading order anywhere else in the equations of motion, continuity, or heat; therefore, we expect it to be a function of the other two non-dimensional parameters in the zonal momentum balance. If a power law functional form is assumed for this relationship, then $V_e/U$ must also be related to $\chi$ through some power law. If it is furthermore assumed that the upper tropospheric eddy velocity $V_e$ is proportional to the near-surface eddy velocity $v_e$ (which is expected, given that the baroclinic kinetic energy primarily resides in the upper and lower troposphere), then the non-dimensional eddy velocity $v_e^*$ should also follow a power law relationship with $\chi$. Indeed, an excellent fit to the simulation data is:
\begin{equation}
\label{eq:v_star_scale}
    v_e^* \sim \chi^{-0.4}.
\end{equation}
The dimensional near-surface eddy velocity may be expressed as:
\begin{equation}
    v_e \sim (fa)^{2/5}U^{3/5}
\end{equation}
The dimensional eddy velocity exhibits a slight dependence on $f$ and $a$ and exhibits a slightly stronger than square root dependence on the dimensional meridional temperature gradient. Ultimately, the dependence on $\chi$ is physically motivated, but the power law fit itself is an empirical result. We speculate the inverse dependence on $\chi$ in (\ref{eq:v_star_scale}) captures a reduction in EKE due to the effects of EMFD across strong zonal jets, whose strength tend to increase with decreasing $\Omega_*$ through thermal wind balance. The strong zonal shears in these stronger jets promote faster conversion rates through EMFD of EKE to mean flow kinetic energy in the upper troposphere \citep{lorenz_available_1955}. This can broadly act to reduce eddy lifetime and dissociate lower troposphere waves from their upper troposphere counterparts, possibly explaining the higher ratios of EKE$_\mathrm{BC}$ to EKE$_\mathrm{BT}$ for lower $\Omega_*$ in Figure \ref{fig:eddy_energies}c. 

\begin{figure}[t]
    \centering
    \includegraphics[width=1.0\linewidth]{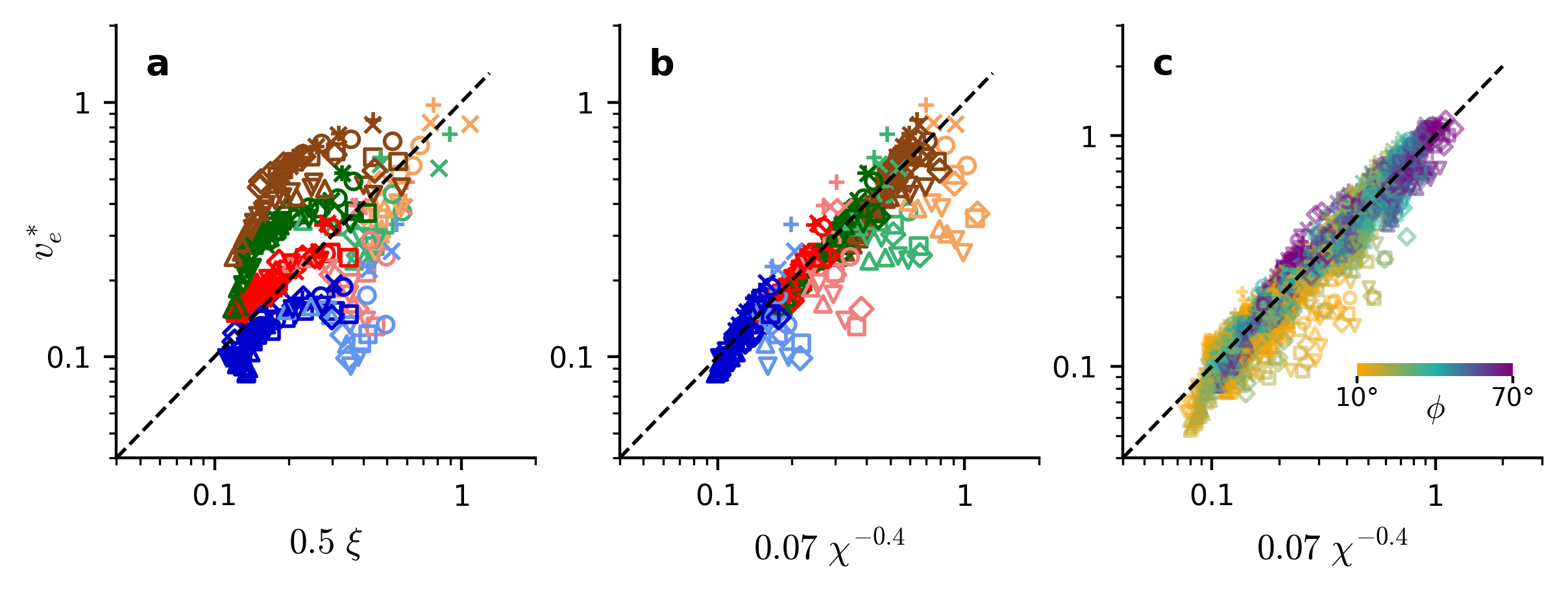}
    \caption{Scalings for the non-dimensional near-surface eddy velocity $v_e^*$ with empirically determined scaling coefficients indicated in the x-axis label. (a) The scaling suggested by \citet{held_scaling_1996} averaged over the baroclinic zones of each simulation. Marker indicators, including color, shape, and shading, are as in Fig. \ref{fig:eddy_energies}. (b) Like a, but using the scaling for $v_e^*$ in (\ref{eq:v_star_scale}). (c) The scaling (\ref{eq:v_star_scale}) applied locally at various latitudes for all simulations. Marker indicators are as in Fig. \ref{fig:lmix}c.}
    \label{fig:v_star}
\end{figure}

The scaling (\ref{eq:v_star_scale}) is able to explain variations in $v_e^*$ within a given rotation rate and across different rotation rates, including 96\% of the $v_e^*$ variance in log-space across all simulations with $\Delta_h \geq 80$~K (Figure \ref{fig:v_star}). It explains 93\% of the variability in local $v_e^*$ in log-space across all simulations with $\Delta_h\geq 80$~K (Fig. \ref{fig:v_star}c). The scaling suggests $v_e^*$ scales with latitude like $f^{4/5}$ (accounting for the $f$-dependence in $U$)---the GCM results corroborate this prediction, as $v_e^*$ generally increases with latitude. The scaling degrades for the lowest $\Delta_h$, $\gamma=0.7$ simulations and locally for some low latitude points. Appendix B explains the cause for this discrepancy: in these weakly forced simulations, the thermal stratification of the troposphere is primarily maintained by convection \citep{schneider_tropopause_2004}, which quickly damps near-surface temperature anomalies before any significant baroclinic instability can ensue. The scalings are only expected to hold in the regime where baroclinic eddy fluxes module the thermal structure of the atmosphere. Note that the low scaling coefficient of 0.07 (shown in Figure \ref{fig:v_star}b) is likely because the near-surface eddy velocity is plotted, but the scaling is intended for the much higher upper-troposphere eddy velocity. 

\subsection{The correlation coefficient}

The correlation coefficient $c$, which appears in the expression (\ref{eq:diff_closure}) for $D$, is considered negligible in many studies or is implicit in the definition of the mixing length. \citet{thompson_scaling_2006} diagnose the correlation coefficient and find it is approximately constant across the vast majority of their 2-layer QG simulations. However, $c$ varies appreciably across our simulations as well as spatially in a given climate state. This variation was noted by \citet{schneider_scaling_2008}, but a quantitative scaling behavior for $c$ has not been proposed. \citet{swanson_lower-tropospheric_1997} estimate near-surface eddy heat fluxes in Earth's mid-latitude storm tracks using reanalysis data and propose a simple stochastic Lagrangian damped-advective model for understanding turbulent heat transport by eddies. In this model, Lagrangian parcels are advected by a stochastically forced velocity field from an initial state with a prescribed mean temperature gradient, velocity variance, and Lagrangian integral decorrelation timescale $\tau_d$. The temperature anomalies of the parcels relative to the environment are damped according to some timescale $\tau_Q$. This model of the turbulent motions of stirred parcels yields the following analytical expression for $c$ consistent with the functional form plotted in their Figure 10:
\begin{equation}
\label{eq:c_stochastic}
    c = \sqrt{\frac{\tau_d}{\tau_d + \tau_Q}}.
\end{equation}
\citet{swanson_dynamics_1997} fit their model parameters to the data and no closed formula is presented in terms of mean flow parameters. However, we can estimate the Lagrangian integral decorrelation timescale using our eddy mixing decorrelation timescale and the velocity variance with our velocity scaling in the previous section. The decorrelation timescale follows from the velocity and mixing length scalings as $\tau_d = \ell_m/v_e$, yielding:
\begin{equation}
\label{eq:tau_d}
    \tau_d = \frac{1/\sqrt{v_e \beta}}{1 + (\alpha U / v_e)^2}.
\end{equation}

\begin{figure}[t]
    \centering
    \includegraphics[width=0.5\linewidth]{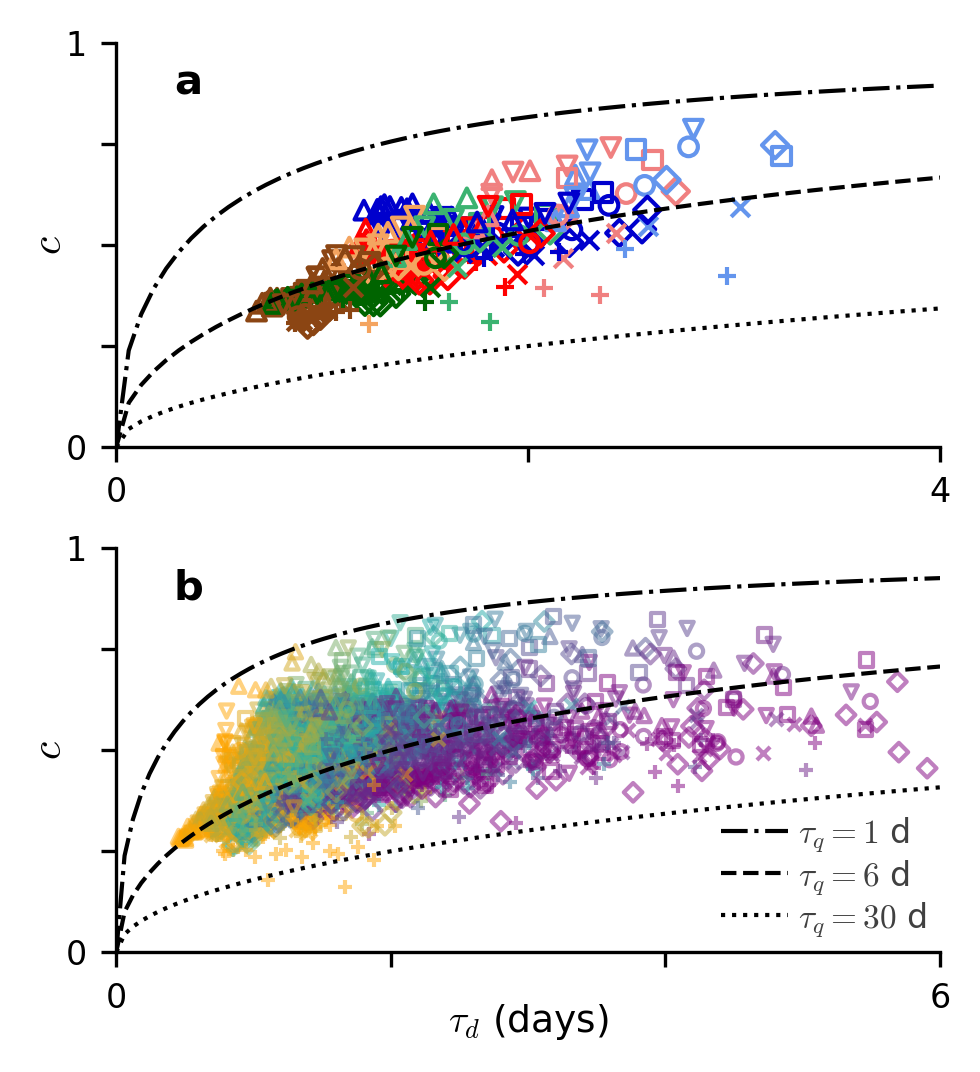}
    \caption{The GCM-calculated correlation coefficient $c$, defined in (\ref{eq:correlation}) plotted against the parameterization for $c$ in (\ref{eq:c_stochastic}). Note that this is a fully closed scaling, with the $v_e^*$ in (\ref{eq:tau_d}) for $\tau_d$ expressed with the scaling (\ref{eq:v_star_scale}).  (a) The quantities averaged over the baroclinic zones of each simulation. Marker indicators are as in Fig. \ref{fig:eddy_energies}. (b) The quantities evaluated locally, with marker indicators as in Fig. \ref{fig:lmix}c.}
    \label{fig:correlation}
\end{figure}

Variations in $\tau_Q$ are likely to depend intimately on model-specific details of convective and radiative processes that might not generalize well to real planetary systems like Earth. On the contrary, variations in $\tau_d$ should be related to dynamical mechanisms general to all planetary systems. Therefore, our proposed scaling for $c$ depends only on $\tau_d$, and we set $\tau_Q=6$~days. Figure \ref{fig:correlation} plots the GCM-calculated correlation coefficient against $\tau_d$ averaged over the baroclinic zones in each simulation in panel a and applied locally in panel b. The plotted lines are the functional relationship for $c$ in (\ref{eq:c_stochastic}) with different values of $\tau_Q$ for each line as indicated in the plot legend. The upper and lower lines represent upper and lower limits based on our convective and radiative relaxation timescales of 1 and 30 days, respectively (note that some simulations have a 60 day relaxation timescale, and consequently tend to have lower $c$). The dashed line indicates the empirically determined best estimate of $\tau_Q=6$ days. 

The scaling (\ref{eq:c_stochastic}) captures the general increase in $c$ with rotation rate observed in the GCM simulations. However, the accuracy of the scaling, especially when applied locally, is limited by the complicating effects of convection, whose sporadic and intense nature cannot be accounted for with a simple relaxation timescale as with radiative processes. In lower latitudes and lower $\Delta_h$ simulations (indicated with the lighter shading in panel a), we expect the points to fall closer to the $\tau_Q=1$~day line due to the increased prevalence of convection, as is indeed the case in \ref{fig:correlation}a and \ref{fig:correlation}b. Also, as expected, simulations with higher $\gamma$ tend to have lower $c$ for a given $\tau_d$, since convection is less prevalent. It is reassuring that maximum and minimum values of $c$ correspond to the theoretical upper and lower bounds (Figure \ref{fig:correlation}b). The complicated interplay between convective and radiative processes will ultimately introduce errors in local predictions of the diffusivity. Fortunately, over-predictions in $v_e$ somewhat compensate under-predictions of $c$, since convection has opposite effects on each quantity.

\subsection{The diffusivity}

Combining the three component scalings of the previous sections for $\ell_m^*$, $v_e^*$, and $c$ in (\ref{eq:mixl_supp_star}), (\ref{eq:v_star_scale}), (\ref{eq:c_stochastic}) yields an expression for the near-surface eddy diffusivity $D$. The non-dimensional diffusivity $D^* = D/U\lambda = c v_e^* \ell_m^*$ is given by:
\begin{equation}
\label{eq:D_star}
    D^* = c\frac{\xi^{1/2}\chi^{-3/5}}{1 + \alpha^2 \chi^{4/5}}.
\end{equation}
The dimensional diffusivity $D$ is given by:
\begin{equation}
\label{eq:D_dimensional}
    D = c \frac{U^{0.9}(fa)^{0.6}\beta^{-0.5}}{1 + \alpha^2 U^{0.8}(fa)^{-0.8}}
\end{equation}
We have expressed (\ref{eq:D_dimensional}) to facilitate direct comparisons with the \citet{held_scaling_1996} scaling that predicts $D\sim (U\xi)^{3/2}\beta^{-1/2}$. The numerator of (\ref{eq:D_dimensional}) is analogous to this, except our eddy velocity scales like $v_e\sim (U\chi^{0.4})^{3/2} = U^{0.9}(fa)^{0.6}$. The scaling (\ref{eq:D_dimensional}) also features an additional term in the denominator accounting for the direct suppressive effects of kinematic mean flow advection on the near-surface mixing length, as in \citet{ferrari_suppression_2010}. Importantly, the dimensional eddy diffusivity depends on $\chi$, not the criticality $\xi$. 

\begin{figure}
    \centering
    \includegraphics[width=1.0\linewidth]{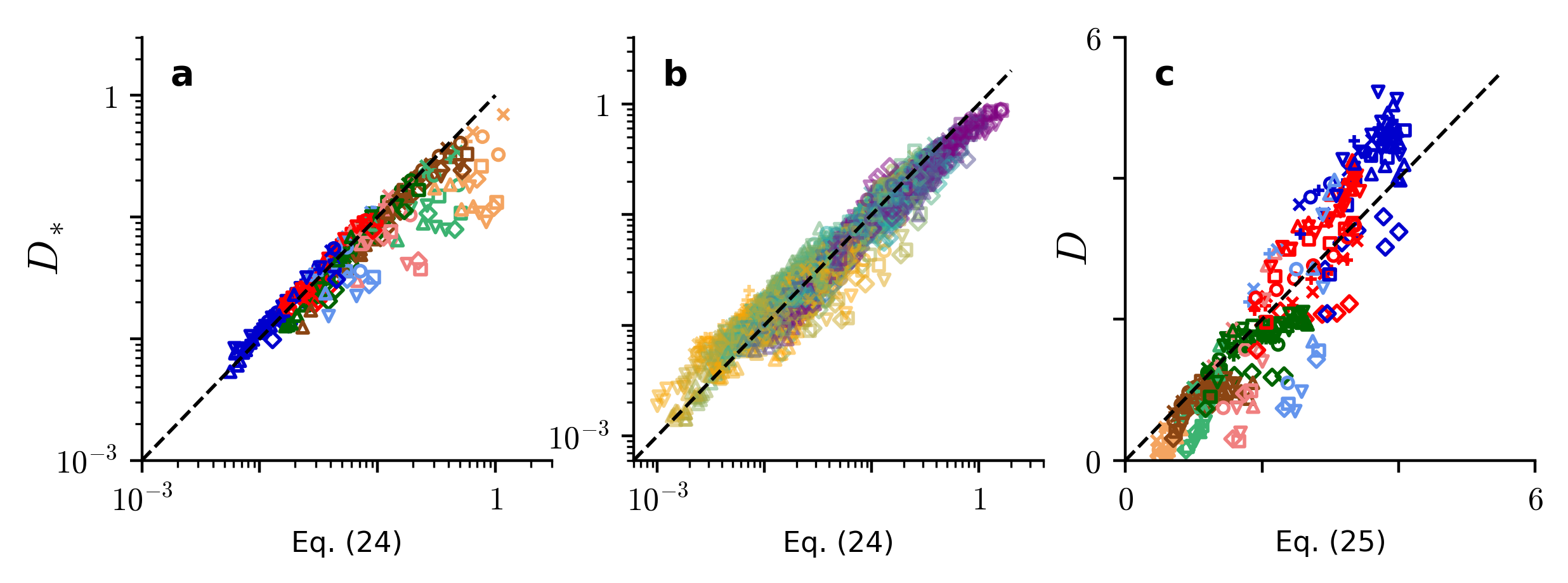}
    \caption{Closed scalings for the non-dimensional near-surface eddy diffusivity against the GCM-calculated values. (a) Non-dimensional $D^*$ scaling, calculated with Eq. (\ref{eq:D_star}), averaged over the baroclinic zones of simulations. Values are shown on a log-log axis. Marker indicators are as in Fig. \ref{fig:eddy_energies}. (b) Like panel a, but evaluated locally at various latitudes from the simulations in panel a. Latitude is indicated with the color shading as in Fig. \ref{fig:lmix}c. (c) Like panel a, but showing the dimensional diffusivity scaling, calculated with Eq. (\ref{eq:D_dimensional}), in units $\SI{e6}{\m\squared\per\s}$.}
    \label{fig:D_scale}
\end{figure}

\begin{figure}
    \centering
    \includegraphics[width=0.4\linewidth]{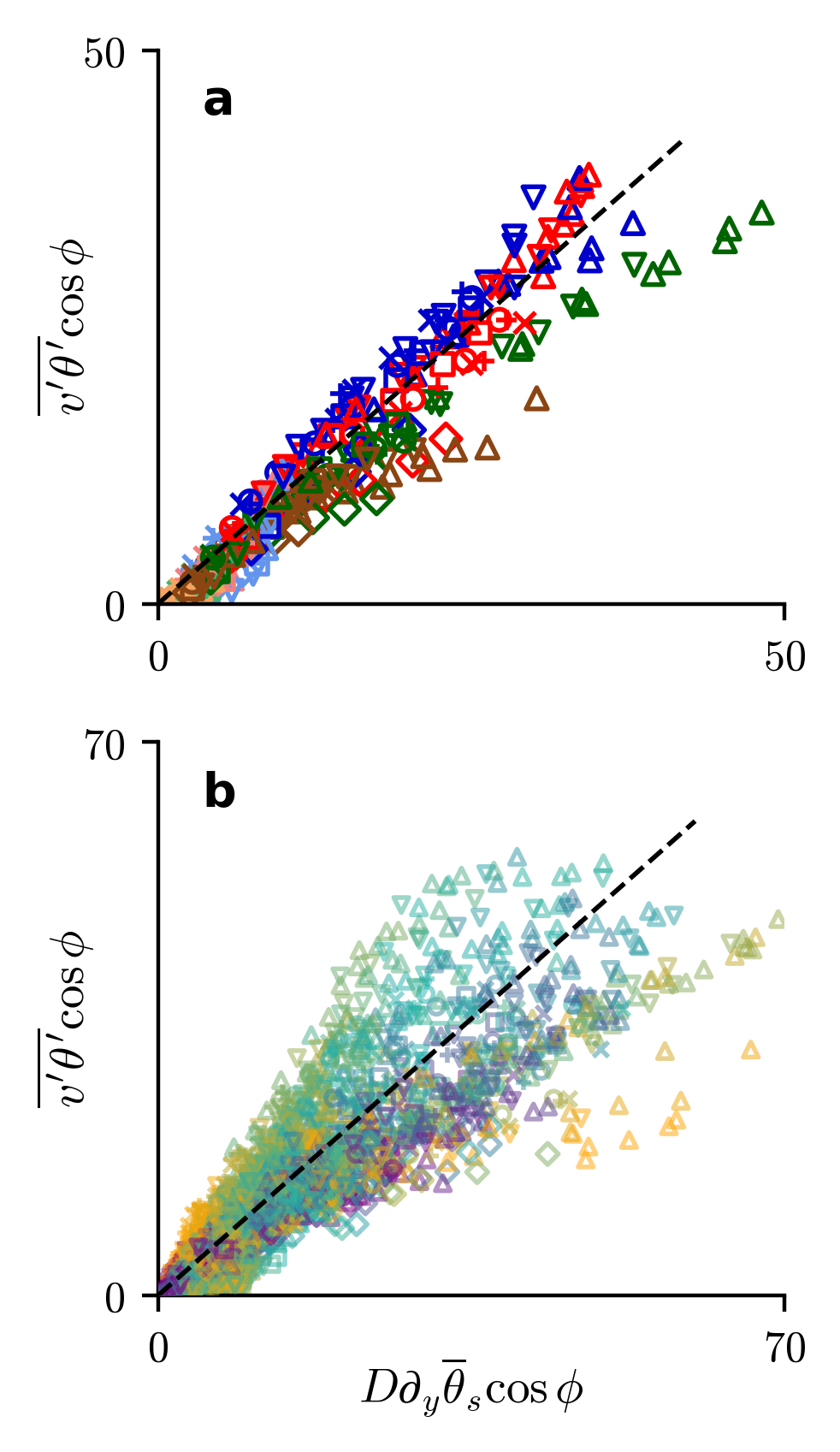}
    \caption{Closed scalings for the dimensional near-surface eddy heat flux against the GCM-calculated values. The scalings are obtained by multiplying the dimensional diffusivity $D$ by $\partial_y\overline{\theta}_s\cos\phi$, with $y = a\phi$. The units for all quantities are $\SI{}{\K\m\per\s}$. (a) Near-surface eddy heat fluxes averaged over the baroclinic zones of the simulations. Marker indicators are as in Fig. \ref{fig:eddy_energies}. (b) Like panel a, but evaluated locally at various latitudes from the simulations in panel a. Latitude is indicated with the color shading as in Fig. \ref{fig:lmix}c.}
    \label{fig:heatflux_scale}
\end{figure}

The scaling (\ref{eq:D_dimensional}) explains 97\% of the variability in $D^*$ in log-space across the $\Delta_h > 60$~K simulations (Figure \ref{fig:D_scale}a). As expected from the velocity scaling results, $D^*$ is over-estimated for the low-$\Delta_h$, low-$\gamma$ simulations where the thermal structure is primarily maintained by convection (see Appendix B). The scaling is also able to capture local variations across three orders of magnitude of $D^*$ (Figure \ref{fig:D_scale}b). Panel c shows the baroclinic zone-averaged dimensional diffusivity predictions, for which it explains 87\% of the variability across all simulations. For both $D^*$ and $D$, the primary source of error outside of the low-$\Delta_h$ simulations is the scaling for $c$. The scaling overestimates $c$ for faster rotation rate simulations, causing a slight over-prediction of $D$ in the faster rotation rate simulations and an under-prediction in some slower rotation rate simulations.

The dimensional near-surface eddy heat flux is the essential quantity that describes eddy effects on the large-scale climate. Figure \ref{fig:heatflux_scale} shows the predictions from the proposed scaling against GCM-calculated values averaged over baroclinic zones in panel a and applied locally in panel b. The scaling predictions are obtained by multiplying the scaling for the dimensional diffusivity $D$ in (\ref{eq:D_dimensional}) by $\partial_y\overline{\theta}_s \cos\phi$. The scaling is able to explain 91\% of the variance in simulation-averaged eddy heat fluxes and 79\% of the variance in local eddy heat fluxes across all simulations. Again, the primary source of  error in $D$ comes from the scaling for $c$, especially in local predictions. We show in section 5 that the diffusivity scaling (\ref{eq:D_dimensional}), when applied as a local parameterization in an EBM, can explain the variability in pole-to-equator temperature contrasts across the GCM simulations.

\section{Applying the scalings in an EBM to explain global climate}

\subsection{The EBM}

We approximate the near-surface energy balance using a simplified diffusive EBM in which the near-surface potential temperature in steady-state is set by a balance between diffusive meridional heat transport and radiative heating. Using the near-surface equilibrium temperature profile $\theta_E$, we can express the time-dependent zonal-mean near-surface potential temperature $\overline{\theta}_s$ as:
\begin{equation}
\label{eq:EBM_balance}
    \fpd{\overline{\theta}_s}{t} = \frac{1}{a^2 \cos\phi}\fpd{}{\phi}\left( D \cos\phi \fpd{\ov{\theta}_s}{\phi}\right) - \frac{\overline{\theta}_s - \theta_E(\phi)}{\tau}.
\end{equation}
This EBM can be run to steady state and compared against the near-surface GCM $\overline{\theta}_s$. In the low-latitude tropical region, the Hadley circulation is the dominant mechanism of heat transport and must be parameterized to obtain any accurate global distribution of temperature. While the mean advection of heat by the Hadley circulation is not a diffusive process, the effect of its heat transport can be approximated diffusively as in \citet{mbengue_linking_2018} with a so-called ``Hadley adjustment." With this adjustment, the full diffusivity $D$ in (\ref{eq:EBM_balance}) is expressed as:
\begin{equation}
\label{eq:EBM_D}
    D = D_e(\phi)[1- w(\phi)] + D_hw(\phi, \phi_h),
\end{equation}
where $w(\phi) \in [0,1]$ is a weighting function that smoothly transitions the dominant heat transport mechanism from the Hadley cell to mid-latitude transient eddies between the tropics and the mid-latitudes. The weighting $w$ is a function of both local latitude $\phi$ and Hadley cell edge latitude $\phi_h$, which is diagnosed from the GCM output. Such Hadley adjustments are common in the literature \citep{lindzen_realistic_1977, siler_insights_2018, armour_meridional_2019} and correct for the distinct dynamical differences between the Hadley circulation and the mid-latitudes with respect to their heat transport behavior. We adopt a similar functional form as in \citet{siler_insights_2018} and \citet{armour_meridional_2019} for the weighting function:
\begin{equation}
\label{eq:ebm_w}
    w(\phi, \phi_h) = \exp\left(-\frac{1}{2}\frac{\phi^2}{(\phi_h/2)^2}\right).
\end{equation}
The weighting function has a Gaussian shape with width $\phi_h/2$. Similar to \citet{mbengue_linking_2018}, $D_h$ is set to a high value, \SI{8e6}{\m\squared\per\s}, to emulate the efficient heat transport and flatter temperature gradients observed in the tropics. While this is a crude treatment of the Hadley cell heat transport, the main goal of the EBM here is to assess the ability of the scalings to reproduce the mid-latitude heat transport by transient eddies. This functional form of the Hadley correction emulates sufficiently accurate Hadley cell transport, which allows us to focus on the effect of our scalings on the atmospheric heat transport poleward of the tropics. For the eddy diffusivity $D_e$, we use the scaling (\ref{eq:D_dimensional}).

\begin{figure}[t]
    \centering
    \includegraphics[width=0.4\linewidth]{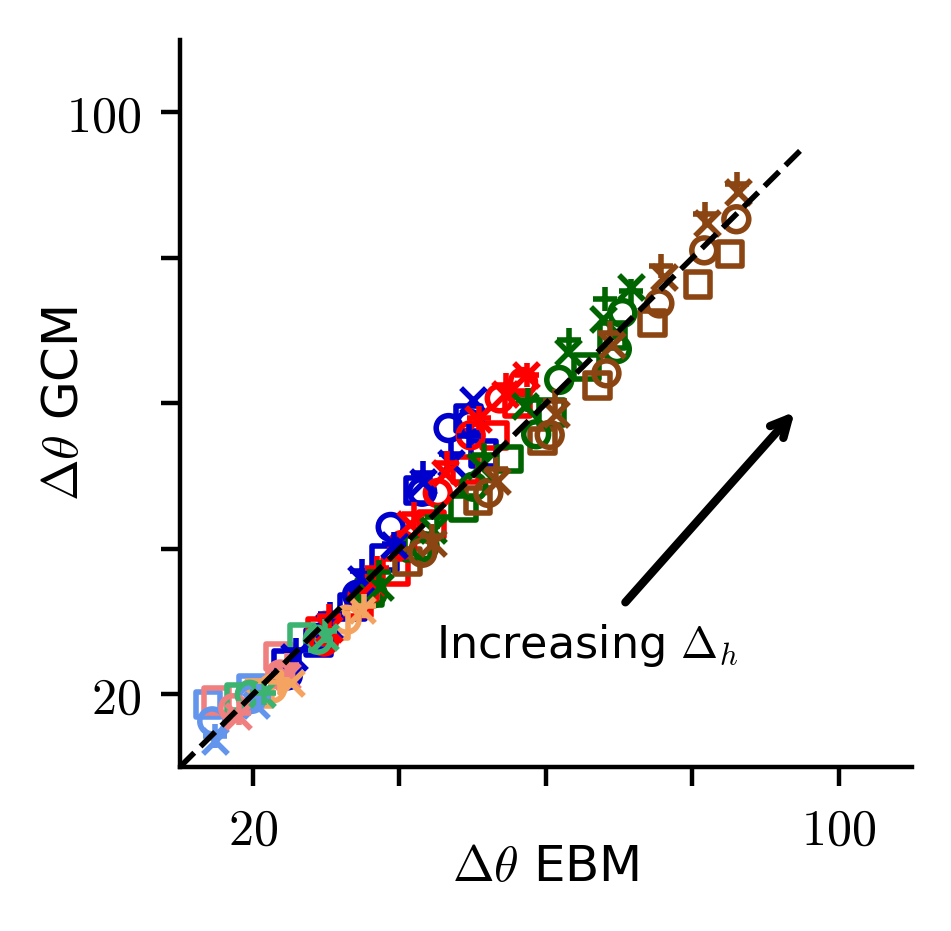}
    \caption{Equator-to-pole $\ov{\theta}_s$ contrasts for the EBM and the GCM for equinox ($\phi_0 = 0\degree$) simulations. Marker indicators are as in Fig. \ref{fig:eddy_energies}. }
    \label{fig:ebm_temp_dif}
\end{figure}

The EBM is run for all equinox ($\phi_0=0\degree$) simulations for every combination of $\Delta_h$ and $\Omega_*$ in Table \ref{tab:param_sweep} for the $\gamma$ values of 0.7, 0.8, 0.9, and 1.0. Since (\ref{eq:D_dimensional}) exhibits no dependence on stratification, the EBM solutions for different $\gamma$ only differ due to differences in $\phi_h$ between the GCM simulations. We do not experiment with the $\phi_0>0\degree$ simulations because upper tropospheric easterlies develop near the equator in the cross-equatorial winter Hadley cells, inducing up-gradient heat transport by the Hadley circulation in the winter hemisphere near the equator. To model these asymmetrically forced climates, a more sophisticated treatment of the Hadley circulation heat transport is necessary.

\subsection{EBM Results}

\begin{figure}
    \centering
    \includegraphics[width=1.0\linewidth]{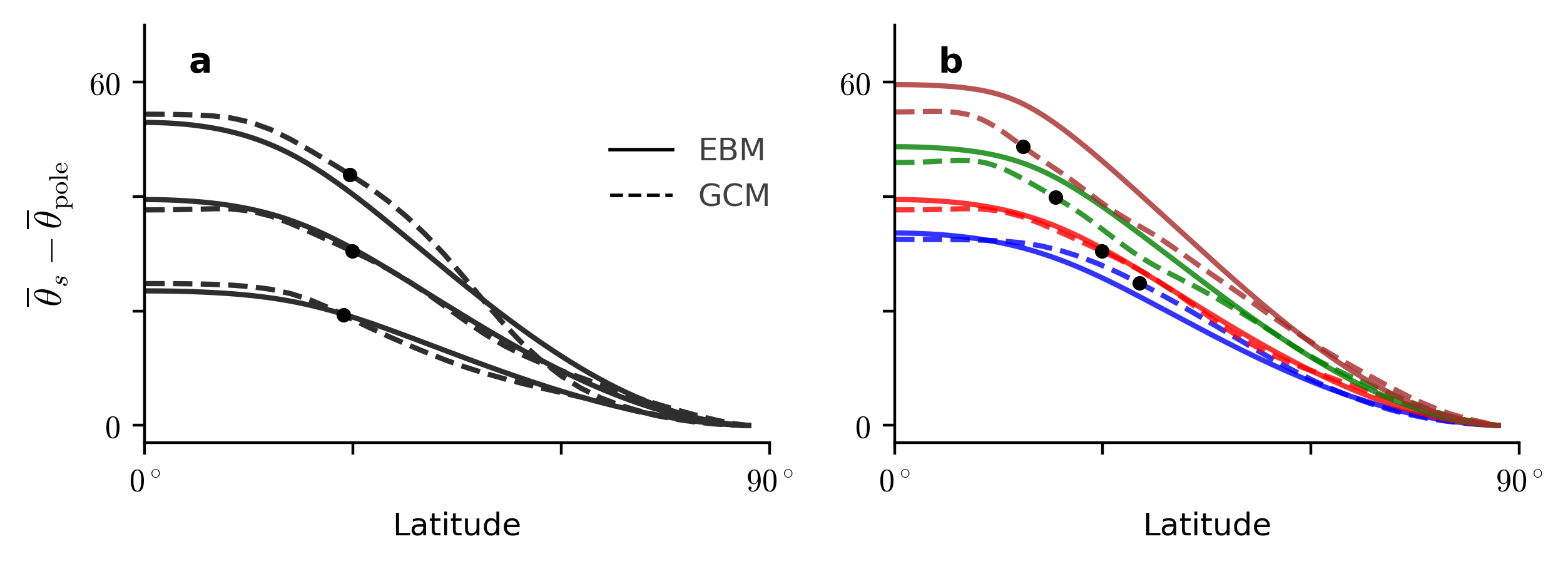}
    \caption{Northern hemisphere temperature profiles for the EBM (solid) and GCM (dashed) near-surface $\overline{\theta}_s$ solutions for different simulations. The temperatures indicated on the y-axis are relative to the corresponding simulation's near-surface polar temperature. The EBM and GCM profiles closest to each other are for the same simulated climate. The black markers on the GCM profile indicate the simulated Hadley cell edge; this latitude corresponds to $\phi_h$ in the EBM Hadley adjustment weighting factor (\ref{eq:ebm_w}). (a) Simulations with $\Delta_h=60~$K (bottom), 120~K (middle), and 180~K (top) for $\phi_0=0\degree$ and $\Omega_*=1$. (b) Simulations with varying rotation rates and $\phi_0=0\degree$ and $\Delta_h=120$~K. The color indicates the rotation rate, as in Fig. \ref{fig:lmix}.}
    \label{fig:ebm_profiles}
\end{figure}

The EBM accurately reproduces the modeled global temperature contrasts, and can accurately reproduce the observed behaviors for varying $\Omega_*$ and varying $\Delta_h$, with some slight biases with rotation rate due to biases in the scaling for $c$ (Figure \ref{fig:ebm_temp_dif}). The EBM captures the rate of increase in pole-to-equator temperature contrast with increasing $\Delta_h$ as well as the faster increase of the pole-to-equator temperature gradient with $\Delta_h$ at faster rotation rates than at slower rotation rates. Slight discrepancies with varying $\gamma$ are primarily due to changes in the Hadley circulation heat transport. The Hadley circulation heat transport generally weakens with increasing $\gamma$ \citep[see Fig. 4a in][]{walker_eddy_2006}. The $D_h$ chosen to emulate the Hadley circulation heat transport was chosen to optimize the $\gamma=0.7$ simulations. Therefore, using a constant $D_h$ for all simulations leads to an over-estimation of Hadley circulation heat transport for higher $\gamma$ simulations, thus causing an under-estimation of the pole-to-equator temperature gradient in the EBM.  

The EBM reproduces well the overall structure of the modeled $\overline{\theta}_s$ profiles. Examples for $\Omega_*=1$ and $\Delta_h$ varying between 60, 120, and 180 K are shown in Figure \ref{fig:ebm_profiles}a, and examples for $\Delta_h=120$~K simulations for four different rotation rates are shown in Figure \ref{fig:ebm_profiles}b. For the more strongly forced high $\Delta_h$ simulations, advection by the mean circulation, including the Hadley, Ferrel, and polar cells, is stronger, causing stronger variations in the slope of the temperature profile with latitude. A simple diffusive EBM cannot capture such variations, but it still captures the global equator-to-pole temperature contrast.

\section{Discussion}

\citet{held_macroturbulence_1999} discusses how an understanding of the near-surface eddy heat flux can form the basis of a general circulation theory. The zonally averaged meridional mass transport in isentropic coordinates can be approximated in terms of the near-surface eddy heat flux as \citep[see][]{andrews_planetary_1976, held_surface_1999}:
\begin{equation}
    V \approx \frac{\ov{v'\theta'}^s}{{g\overline{(\partial_p \theta)}^s}}.
\end{equation}
The diffusive closure for the eddy heat flux developed in section 3 can be used to predict $\ov{v'\theta'}^s$, and therefore $V$, as a function of the mean flow. What remains is an a priori prediction of the vertical stratification of the extratropical troposphere, which may consist of either a prediction of the lapse rate or the tropopause height. \citet{schneider_tropopause_2004} and \citet{schneider_scaling_2008} propose one such explanation, namely that atmospheres tend to organize themselves into an O(1) criticality state across a broad range of atmospheres. However, it remains unclear how general this relation should be \citep{jansen_macroturbulent_2012, jansen_vertical_2013}. An alternative route to a complete understanding of the mid-latitude circulation might be to predict the macroturbulent transport of PV in the upper troposphere directly through some diffusive closure, which is also challenging since eddy PV fluxes more strongly distort the mean PV gradient than near-surface eddy heat fluxes distort the temperature gradient \citep{held_macroturbulence_1999}. The results of this paper suggest that kinematic effects may be important in our understanding of these upper tropospheric eddy fluxes. Also based on the scaling analysis in section 3, the parameter $\chi$ should be relevant for the upper troposphere PV fluxes, too. In addition to a complete understanding of the mid-latitude circulation, a closed theory of the tropical Hadley circulation and how it interacts with eddies of mid-latitude origin will be necessary for a closed general circulation theory.

While this paper focuses primarily on dry dynamics, with the stabilizing effect of latent heat release encoded through the dry convection scheme, the scalings should be relevant for describing moist atmospheres, such as Earth. The scaling for the dimensional diffusivity in section 3 exhibits no dependence on the stratification, and the GCM simulations with varied convective lapse rates corroborate this scaling prediction. This suggests that the effects of moisture on tropospheric stability may not be important for near-surface eddy heat fluxes in Earth-like climates. Indeed, \citet{ogorman_energy_2008} find that EKE in the baroclinic zones of idealized moist atmospheres scales with the dry MAPE, suggesting that the dry dynamics might generalize well to moist atmospheres. However, as the effects of convection in the dry simulations indicate, the scalings might become erroneous in regions and climates where convection dominates the maintenance of the vertical stratification. Future work will further examine the relevance of the proposed scalings in moist atmospheres.

\section{Conclusions}

We present simulation results of near-surface eddy heat fluxes from 323 idealized GCM simulations across a wide parameter regime of pole-to-equator temperature gradients, rotation rates, seasonality, and vertical stratification. We propose a new scaling theory for the near-surface macroturbulent heat transport that accurately reproduces the simulated heat flux, even in parameter regimes where classical 2-layer QG theories fail. There are two key distinctions between our proposed theory and other classical theories, such as that proposed by \citet{held_scaling_1996}:

\begin{enumerate}
    \item We incorporate the kinematic effect of mean flow suppression arising from a difference in eddy phase speeds and the near-surface mean zonal flow \citep{ferrari_suppression_2010}. When this mean flow suppression is weak, the mixing length scaling reduces to the Rhines scale, as in \citet{held_scaling_1996}. However, for the majority of simulations, this kinematic reduction of the mixing length cannot be ignored to obtain accurate scalings.
    
    \item Our eddy velocity scales with $U\chi^{0.4}$ instead of $U\xi$. The parameter $\chi$ is a planetary scale Rossby number comparing the strength of the thermal wind shear to the planetary velocity scale $fa$. A power law relationship between the eddy velocity and $\chi$ is suggested from non-dimensionalizing the zonal momentum balance for the upper troposphere and examining the leading order balance. This scaling implicitly accounts for non-linear interactions between eddies and the mean zonal jet (through eddy momentum fluxes), a process neglected in homogeneously forced QG theories.
\end{enumerate} 
Classical QG theories predict that the near-surface eddy heat flux diffusivity scales cubically with the temperature gradient \citep{held_scaling_1996}. The scalings developed in this work indicate that the diffusivity depends nearly linearly on the temperature gradient in the limit of weak suppression and exhibits nearly zero dependence on the temperature gradient in the limit of strong suppression. Earth-like climates reside between these two limits. The scalings (\ref{eq:D_star}) and (\ref{eq:D_dimensional}) also feature a latitudinal dependence which is absent in other theories. The scalings do not exhibit any dependence on stratification, so we expect they should apply to moist atmospheres, such as Earth's. The complicated dependence of the eddy diffusivity on the thermal wind shear and latitude might also explain phenomena in Earth's storm tracks, such as the midwinter suppression in EKE noted in the Pacific storm track region \citep{nakamura_midwinter_1992, schemm_eddy_2018, hadas_suppression_2021}.

The proposed scalings can be used as a diffusive closure in a simple EBM to understand the global distribution of surface temperatures across different climate states (Section 5). Integrating a realistic treatment of Hadley circulation heat transport into the EBM would provide a simple, fully closed 1D model that could explain the global distribution of temperature, as well as interactions between the zonally averaged extratropical and the Hadley circulations. Altogether, the proposed scalings can form the basis of a general circulation theory, upon which an understanding of the thermal stratification of the extratropical troposphere and the dynamics of the Hadley circulation may be built. 

\acknowledgments This material is based upon work supported by the National Science Foundation Graduate Research Fellowship under Grant No. 2139433. R.E. acknowledges very helpful discussions with Raffaele Ferrari for the mixing length formulation.

\datastatement The idealized GCM used is publicly available and scripts for generating the simulation data will be made publicly available on Github after manuscript acceptance.


\appendix[A]
\label{section:v_equipartition}
\appendixtitle{The relationship between the eddy velocity and mixing length}

\citet{schneider_scaling_2008} approximate the bulk EAPE integrated over a baroclinic zone using a Taylor series expansion over the meridional width of the baroclinic zone temperature profile. They then assume this EAPE is equipartitioned with EKE. However, for a local closure, this assumption can lead to errors (evidenced by local mixing lengths that are substantially smaller than the deformation radii, and therefore the baroclinic zone width, in many of the simulations in Fig. \ref{fig:lmix}c). Instead, we may use a local definition of MAPE in terms of our locally defined mixing length to estimate the EAPE and assume it is proportional to the local EKE at a given latitude. Doing so will yield a statement of equipartitioning between the local eddy velocity scale $v_e$ and the local eddy mixing length $\ell_m$. To see this, first consider Lorenz's original approximate expression for EAPE in terms of potential temperature variance on pressure surfaces \citep{lorenz_available_1955}. Similar to \citet{schneider_scaling_2008}, we estimate the vertically averaged quantity in terms of its near-surface value. The vertically and zonally averaged EAPE per unit mass is then approximated as (with units $\SI{}{\m\squared\per\s\squared}$):
\begin{equation}
\label{eq:eape}
    \mathrm{EAPE} \approx -\frac{c_p}{2} \frac{\kappa}{p_s} \left(\frac{p_s}{p_0}\right)^{\kappa}\frac{\overline{\theta_s'^2}}{\overline{\partial_p\theta}^{s}}.
\end{equation}
The heat capacity at constant pressure is denoted by $c_p$ and $\kappa=R_d/c_p$. The reference pressure $p_0 = 1000~\mathrm{hPa}$ is approximately equal to the near-surface pressure $p_s$, so their ratio is neglected in what follows. The potential temperature variance may be exactly expressed in terms of the temperature gradient using our definition of the mixing length in Eq. (\ref{eq:mixl_definition}). Substituting (\ref{eq:mixl_definition}) into (\ref{eq:eape}) gives:
\begin{equation}
\label{eq:eape_mixl}
    \mathrm{EAPE} \approx -\frac{c_p}{2} \frac{\kappa}{p_s} \frac{\left(a^{-1}\partial_\phi \overline{\theta}_s\right)^2 \ell_m^2}{\overline{\partial_p\theta}^{s}}.
\end{equation}
Converting the vertical derivative from pressure to height coordinates using the hydrostatic relation, expressing this gradient in terms of the deformation radius (\ref{eq:rossby_outer}), and utilizing the ideal gas law, the EAPE can be succinctly expressed in terms of $U$ by:
\begin{equation}
    \mathrm{EAPE} \approx U^2 \frac{\ell_m^2}{\lambda^2}.
\end{equation}
This relation asserts that the EAPE is proportional to the MAPE over the characteristic mixing length of eddies at a given latitude. Assuming equipartitioning between EKE and EAPE furthermore yields:
\begin{equation}
\label{eq:v_scale_simple}
    v_e^2 \sim U^2  \frac{\ell_m^2}{\lambda^2},
\end{equation}
where we assumed that the near-surface EKE is proportional to the total EKE (this is an accurate assumption for our simulations). This can be rewritten in non-dimensional terms as $v_e^* \sim \ell_m^*$. This is exactly the relation derived in \citet{held_scaling_1996} using inverse energy cascade arguments. If $\ell_m^*$ scales like $\xi$, then we recover the result of \citet{held_scaling_1996}. Unfortunately, the scaling for $\ell_m^*$ (\ref{eq:mixl_supp_star}) itself depends on $v_e^*$ in a complicated way through the mean flow suppression, so it cannot simply be plugged into (\ref{eq:v_scale_simple}) to obtain a closed relation for $v_e^*$ as in \citet{held_scaling_1996}. Additionally, the eddy velocity scale will depend not only on the mixing length near the surface, but also the vertical structure and extent of the eddy heat flux. While the simulation results indicate that $v_e^*$ and $\ell_m^*$ are qualitatively similar, their proportionality has quantitative errors, particularly between different rotation rates. Instead, in the main text, we use scaling arguments in section 3b to determine the eddy velocity scale $v_e$.  

\appendix[B]
\label{section:convection}
\appendixtitle{Convective interference with the eddy velocity scaling in weakly forced climates}

The inaccuracies apparent for the $\gamma=0.7$ simulations with $\Delta_h\leq 60~$K can be explained by the effects of convection. Consider that if the atmosphere were in a state of radiative-convective equilibrium, there would be no meridional circulation and therefore no horizontal EKE to maintain the principal zonal momentum balance assumed in section 3b. Generally, the scaling (\ref{eq:v_star_scale}) should only be valid if radiative heating and cooling are predominantly balanced by eddy heat fluxes. These heat fluxes act as a source of wave activity that propagates into the upper troposphere and induces the upper tropospheric EMFD whose zonal momentum tendency must be balanced by the Eulerian-mean Ferrel circulation. For low-$\Delta_h$ simulations, we expect weak temperature gradients, and therefore weaker eddy heat fluxes due to the lower EKE suggested from the scaling in (\ref{eq:v_star_scale}). To stabilize the vertically unstable RE profile in the extratropics, the dry convection scheme compensates for the weak eddy activity to maintain a statically stable vertical thermal structure. This transition between eddy-dominated and convection-dominated maintenance of the thermal stratification of the troposphere for weakly forced climates has been identified by \citet{schneider_tropopause_2004} and \citet{schneider_self-organization_2006}.

\begin{figure}[h]
    \centering
    \includegraphics[width=0.45\linewidth]{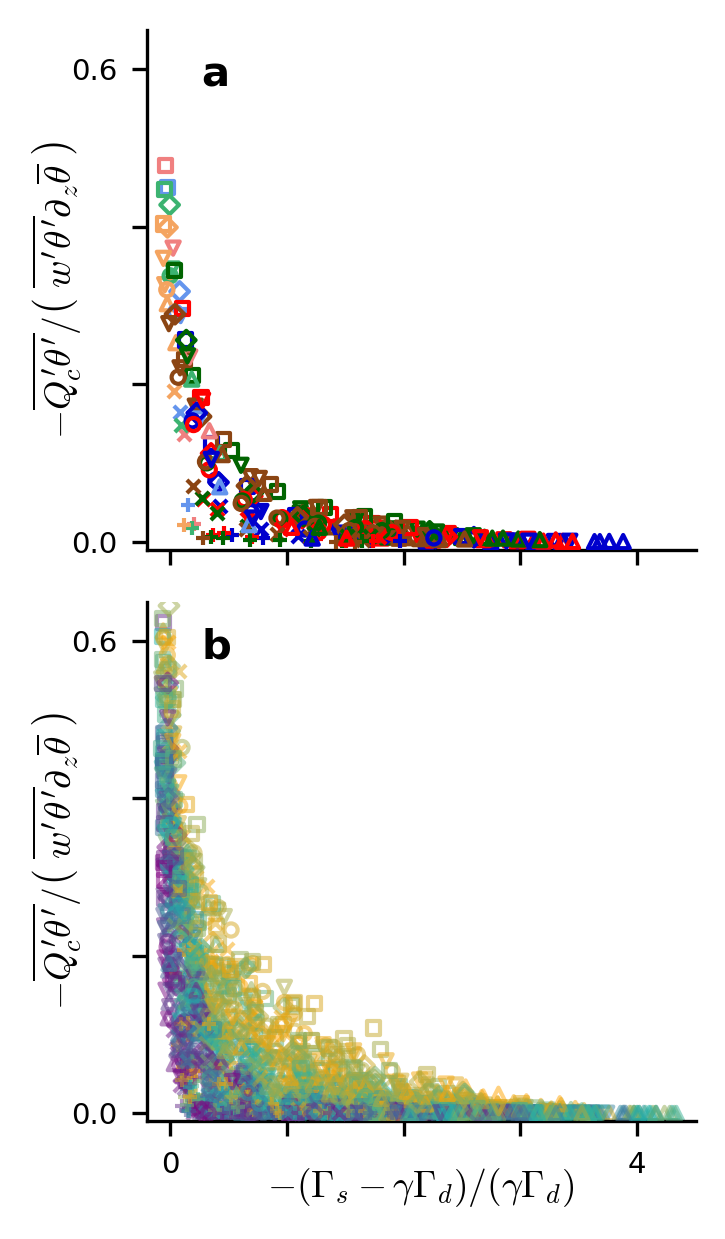}
    \caption{The ratio of the rate of destruction of potential temperature variance to the rate of conversion of potential temperature variance to kinetic energy plotted against the normalized difference between the near-surface lapse rate $\Gamma_s$ and the convectively neutral lapse rate $\gamma\Gamma_d$. (a) Baroclinic-zone averaged values, with marker indicators as in Fig. \ref{fig:eddy_energies} and (b) applied locally with marker indicators as in Fig. \ref{fig:lmix}. }
    \label{fig:lapserate_thetavar}
\end{figure}

Analysis of the potential temperature variance budget, which is closely related to the EAPE budget, confirms the interference of convection in the eddy energy budget for weakly forced simulations. The steady-state potential temperature variance budget, averaged over a baroclinic zone such that the transport terms can be neglected, is:
\begin{equation}
    \ov{v'\theta'}\frac{1}{a}\fpd{\ov{\theta}}{\phi} = - \overline{w'\theta'} \fpd{\ov{\theta}}{z} + \ov{Q'\theta'}.
\end{equation}
The first term on the left is the rate of production of $\theta$ variance, and the two terms on the right are the conversion from $\theta$ variance to EKE and destruction by diabatic processes, either through radiative ($Q_r$) or convective ($Q_c$) processes. Time- and zonal-anomalous diabatic heating is denoted by $Q'$. The atmospheric thermal structure will be eddy-controlled if generation of temperature variance is largely balanced by baroclinic conversion to EKE; it will be convection-controlled if temperature variance generation is balanced by diabatic destruction.  Figure \ref{fig:lapserate_thetavar}a plots the ratio of destruction of temperature variance by convection to the rate of conversion of temperature variance to velocity variance on the y-axis against the near-surface lapse rate $\Gamma_s$ relative to the convectively neutral lapse rate $\gamma\Gamma_d$ for each simulation. 

For many low-$\Delta_h$ simulations, convection destroys a large fraction of the produced temperature variance, thus limiting the baroclinic conversion of potential to kinetic energy (Figure \ref{fig:lapserate_thetavar}a). This fractional rate of destruction rapidly increases when the mean state near-surface lapse rate approaches a convectively neutral state ($\Gamma_s \approx \gamma\Gamma_d$). This ratio of destruction to baroclinic conversion can be represented by a simple power law in terms of $\Gamma_s$ and $\gamma\Gamma_d$. In principle, one might be able to correct for this convective interference with a scaling predicting the diabatic destruction of temperature variance given the near-surface lapse rate. However, this calculation is beyond the scope of this work. The scaling in (\ref{eq:v_star_scale}) overestimates $v_e$, and therefore $D$, for the simulations with $\Delta_h\leq 60$~K and $\gamma=0.7$. There is a strong latitudinal dependence indicating that convective damping is more prevalent at lower latitudes (Figure \ref{fig:lapserate_thetavar}b); this is expected given that near-surface heating rates are stronger at lower latitudes.

%



\newpage
\bibliographystyle{ametsocV6}
\bibliography{references}

\end{document}